\documentclass[aps,twocolumn]{revtex4-1}
\usepackage{dcolumn}
\usepackage{graphicx}
\usepackage{amsmath}
\usepackage{amssymb}
\usepackage{enumitem}
\usepackage{ulem}
\usepackage{verbatim} 
\usepackage{xcolor}

\usepackage{hyperref}
\hypersetup{
    colorlinks=true,
    linkcolor=blue,
    filecolor=blue,      
    urlcolor=blue,
    citecolor = blue
}

\newcommand{\yambo}{\textsc{yambo}}

\newcommand{\editor}[2]{%
  \expandafter\newcommand\csname #1note\endcsname[1]{%
    \textcolor{#2}{(\textbf{#1:} \textit{##1})}}%
  \expandafter\newcommand\csname #1\endcsname[1]{%
    \textcolor{#2}{##1}}%
  \expandafter\newcommand\csname #1cancel\endcsname[1]{%
    \textcolor{#2}{\sout{##1}}}%
  \expandafter\newcommand\csname #1change\endcsname[2]{%
    \textcolor{#2}{\sout{##1} ##2}}%
  \newenvironment{#1text}{\color{#2}}{\color{black}}
}
\definecolor{Blu}{rgb}{0.00,0.00,1.00}
\definecolor{Red}{rgb}{1.00,0.00,0.00}
\definecolor{Cyan}{rgb}{0.00,0.50,0.50}
\definecolor{Green}{rgb}{0.00,0.70,0.00}
\editor{DV}{blue}
\editor{AF}{Red}
\editor{CC}{Cyan}
\editor{DAL}{Green}
\renewcommand{\emph}{\textit}
\newcommand{\suppinfo}{Supplemental Material~\cite{supp-info}}

\begin{document}

\title{Efficient full frequency \texorpdfstring{$GW$}{TEXT} for metals using a multipole approach for the dielectric screening}
\author{Dario A. Leon$^{1,2,3}$}
\email{dario.alejandro.leon.valido@nmbu.no}
\author{Andrea Ferretti$^2$}
\author{Daniele Varsano$^{2}$}
\author{Elisa Molinari$^{1,2}$}
\author{Claudia Cardoso$^{2}$}
\affiliation{$^1$ FIM Department, University of Modena \& Reggio Emilia, 41125, Modena (Italy)}
\affiliation{$^2$S3 Centre, Istituto Nanoscienze, CNR, 41125, Modena (Italy)}
\affiliation{$^3$Department of Mechanical Engineering and Technology Management, Norwegian University of Life Sciences, 1430, Ås (Norway)}

\begin{abstract}

The properties of metallic systems with important and structured excitations at low energies, such as Cu, are challenging to describe with simple models like the plasmon pole approximation (PPA), and more accurate and sometimes prohibitive full frequency approaches are usually required. In this paper we propose a numerical approach to $GW$ calculations
on metals that takes into account the frequency dependence of  the screening via the multipole approximation (MPA), an accurate and efficient alternative to  current full-frequency methods that was recently developed and validated for semiconductors and overcomes several limitations of PPA.
We now demonstrate that MPA can be successfully extended to metallic systems by optimizing the frequency sampling for this class of materials and introducing a simple method to include the $\mathbf{q}\to 0$ limit of the intra-band contributions. The good agreement between MPA and full frequency results for the calculations of quasi-particle energies, polarizability, self-energy and spectral functions in different metallic systems confirms the accuracy and computational efficiency of the method. Finally, we discuss the physical interpretation of the MPA poles through a comparison with experimental electron energy loss spectra for Cu. 

\end{abstract}

\maketitle


\section{Introduction}

Many-body perturbation theory provides accurate methods to study the spectroscopic properties of condensed matter systems from first principles~\cite{Onida2002RMP,martin2016book,Marzari2021NatureMat}. Calculations often adopt the so-called $GW$ approximation~\cite{Hedin1965PR,Strinati_1982,Aryasetiawan1998RPP,martin2016book,Reining2018wcms,Golze2019FrontChem}, for which the frequency integration in the evaluation of the self-energy is crucial to the deployment of the method.
The frequency dependence of the screened potential, $W$, is often described within the  plasmon  pole approximation (PPA)~\cite{Hybertsen1986PRB,Zhang1989PRB,Godby1989PRL,vonderLinden1988PRB,Engel1993PRB,larson2013role},  successfully applied to the calculation of quasi-particle energies of semiconductors~\cite{Hybertsen1986PRB}, the homogeneous electron gas~\cite{Hedin1967IJQC}
and simple metals as Al and Na~\cite{Northrup1987PRL,Surh1988PRB,Northrup1989PRB,Cazzaniga2012PRB}, especially for quasi-particles with energies close to the Fermi level. However, the description of the self-energy and the spectral functions for the whole range of frequencies is still challenging and requires expensive full frequency (FF) approaches. 

Despite its success, the use of PPA is problematic  when complex metals are concerned, even for the calculations of quasi-particle energies~\cite{Aryasetiawan1998RPP}. Its applicability for transition and noble metals has often been disputed~\cite{Aryasetiawan1998RPP,Marini2002PRL}, since the approximation is based on the homogeneous electron gas, for which PPA becomes exact in the long wave-length limit~{~\cite{Fetter-Walecka1971book,Hedin1965PR,Giuliani-Vignale2005book}, while it is in principle not strictly valid in the presence of strongly localized $d$-bands. In fact, these metals present complex screening effects due to collective excitations~\cite{Nilsson1983PRB,book_Palik1985}, which result in highly structured energy-loss spectra whose description is unattainable with a single plasmon peak~\cite{book_Palik1985}. Moreover, metals with relevant excitations at low energies, such as Cu,  require a specially accurate description of the low frequency regime, which makes it difficult to determine the PPA parameters since it requires sampling the polarizability at zero frequency~\cite{Marini2002PRL}.

In this context, we have recently developed a multipole approach (MPA) that naturally bridges from PPA to FF treatments of the $GW$ self-energy~\cite{Leon2021PRB}. The method has been implemented in the {\yambo} 
code~\cite{Marini2009CPC, Sangalli2019JPCM} and was validated for bulk semiconductors. We have shown that, for semiconductors, MPA attains an accuracy comparable to that of FF methods at a much lower computational cost, while also circumventing several of the PPA shortcomings.
Here we extend the assessment of MPA validity and performance to the case of metals. We do so by computing quasi-particle energies, together with the full frequency dependence of the self-energy and the spectral function. The approach is similar to the one used for semiconductors~\cite{Leon2021PRB}, with only slight changes in the frequency sampling strategy used in the multipole interpolation.
In the following, we show that MPA is accurate for metallic systems, even in cases in which the use of PPA is challenging.
In addition to MPA, we also propose a simple {\it ab-initio} method to include  intra-band contributions~\cite{Maksimov1988JPEMP,Methfessel1989PRB,Lee1994PRB,Marini2001PRB,Cazzaniga2008PRB} to the dielectric function in the $\mathbf{q} \to 0$ limit, absent in semiconductors. Despite its virtually zero computational cost, it significantly accelerates the convergence of quasi-particle energies with respect to the $\mathbf{k}$-points grid, in systems where the intra-band contributions are dominant.

The paper is organized as follows:
In Sec.~\ref{sec_methods}, we briefly summarize the $GW$ approximation and the MPA approach. In the same Section, we further extend the strategy used in the frequency sampling for the multipole interpolation, with respect to the MPA implementation presented in Ref.~\cite{Leon2021PRB} for semiconductors. We also discuss the relevance of the inclusion of the intra-band contribution to the dielectric function in the limit $\mathbf{q} \to 0$. In Sec.~\ref{sec:results} we first present MPA calculations for simple metals and propose a simple way of including the aforementioned intra-band limit. We then describe in detail the results obtained for Cu, a prototype challenging system for PPA. Finally, in Sec.~\ref{sec:conclusions} we summarize and discuss the main conclusions of this work.
\section{Methods}
\label{sec_methods}
\subsection{Quasi-particle energies within $GW$}

We adopt the $GW$ approximation~\cite{Hedin1965PR,Aryasetiawan1998RPP,martin2016book,Reining2018wcms,Golze2019FrontChem,Strinati_1982} for the evaluation of the electron-electron self-energy, which is computed via a frequency convolution of the one-particle Green's function $G(\omega)$ and the dynamical screened interaction potential $W(\omega)$:
\begin{equation}
    \Sigma^{GW}(\omega) = 
     \frac{i}{2 \pi}\int_{-\infty}^{+\infty} 
     d\omega' e^{-i\omega'\eta}G(\omega-\omega') W(\omega').
    \label{eq:GW}
\end{equation}
In the present work we limit ourselves to the $G_0W_0$ approximation,
although MPA, the method we want to discuss here, can be exploited also within more advanced approaches such as different self-consistent $GW$ schemes~\cite{vanSchilfgaarde2006PRL,Kotani2007PRB,Shishkin2007PRL,Kutepov2012PRB,Kutepov2017CPC,Grumet2018PRB,Friedrichnano2022},
or methods including vertex-corrections~\cite{Shirley1996PRB,Shishkin2007PRL,Chen-Pasquarello2015PRB,Ren2015PRB,Maggio2017JTCT} and cumulant expansions~\cite{Guzzo2011PRL}. 
A more comprehensive discussion of these aspects can be found e.g. in Refs.~\cite{Reining2018wcms,Golze2019FrontChem}.
The present implementation uses as a starting point single-particle energies and wavefunctions computed within Kohn-Sham (KS) density funtional theory (DFT) to then build the non-interacting single-particle Green's function $G_0(\omega)$ and the irreducible polarizability, $X_0(\omega)$.

The dressed polarizability, $X(\omega)$, and the screened interaction, $W(\omega)$, are then numerically evaluated by solving the Dyson equation for each given frequency:
\begin{eqnarray}
     \label{eq:W}
     X(\omega) &=& X_0(\omega) + X_0(\omega) v X(\omega) \\
     W(\omega) &=& \varepsilon^{-1}(\omega) v = v + v X(\omega) v, 
     \nonumber
\end{eqnarray}
where $v$ is the bare Coulomb potential, $\varepsilon$ the dielectric function and, for simplicity, we have omitted the spatial, non-local, degrees of freedom. All the quantities have to be thought as frequency dependent operators or matrices of the form $X(\omega) = X(\mathbf{r},\mathbf{r}',\omega)$, or, when using a plane-wave basis set, $X_{\mathbf{G}\mathbf{G}'}(\mathbf{q},\omega)$.
The quasi-particle (QP) energies $\epsilon_m^{\text{QP}}$ are then computed either by numerically solving the exact QP equation,
\begin{equation}
   \epsilon_m^{\text{QP}} = \epsilon_m^{\text{KS}} + \langle \psi_m^{\text{KS}}|\Sigma(\epsilon_m^{\text{QP}})-v_{xc}^{\text{KS}}|\psi_m^{\text{KS}} \rangle
   \label{qp_ks},
\end{equation}
or its linearized form:
\begin{equation}
   \epsilon_m^{\text{QP}} \approx \epsilon_m^{\text{KS}} + Z_m \langle \psi^{\text{KS}}|\Sigma(\epsilon_m^{\text{KS}})-v_{xc}^{\text{KS}}|\psi_m^{\text{KS}} \rangle,
   \label{qp_ks_l}
\end{equation}
with the renormalization factors $Z_m$ given by
\begin{equation}
   Z_m = \left[ 1-\langle \psi_m^{\text{KS}}|\frac{\partial\Sigma(\omega)}{\partial \omega}\bigg|_{\omega=\epsilon_m^{\text{KS}}}|\psi_m^{\text{KS}} \rangle \right]^{-1} .
   \label{qp_z}
\end{equation}

In the above equations we have made reference to the Kohn-Sham eigelvaues and eigenvectors, $\epsilon_{m}^{\text{KS}}$ and $| \psi_m^{\text{KS}}\rangle$, respectively.

A key quantity in the above formulation is the dynamical part of the inverse dielectric function, $Y \equiv \varepsilon^{-1}-I = v X$, which determines the correlation part of $W$, $W_c \equiv W- v= Y v$, and, through Eq.~\eqref{eq:GW}, the correlation part of the self-energy, $\Sigma_c$. With the purpose of avoiding the expensive numerical evaluation of the frequency convolution in $\Sigma_c$, Eq.~\eqref{eq:GW}, as required e.g. by full frequency real axis (FF-RA) approaches~\cite{Marini2002PRL,Huser2013PRB} or contour deformation (FF-CD) techniques~\cite{Godby1988PRB,book_Anisimov2000,Kotani2007PRB}, $Y$ or $X$ have been the target of several analytical simplifications like the plasmon pole approximation (PPA)~\cite{Hybertsen1986PRB,Zhang1989PRB,Godby1989PRL,vonderLinden1988PRB,Engel1993PRB} or the multipole approach (MPA)~\cite{Leon2021PRB}, briefly sketched below.

\subsection{The multipole approach}

The multipole approximation is inspired by the Lehmann representation of the polarizability $X$. At the independent particle level, $X$ (equal to $X_0$) is written in a compact way as a sum of poles with vanishing imaginary part corresponding to all possible single particle transitions (here considered at the Kohn-Sham level for simplicity) of energy $\Omega^{\text{KS}}$ and probability amplitude $R^{\text{KS}}$:
\begin{equation}
    X_0(\omega) = \sum_{n}^{N_T} \frac{2 R^{\text{KS}}_{n} \Omega^{\text{KS}}_{n}}{\omega^2-(\Omega^{\text{KS}}_{n})^2},
    \label{eq:X0}
\end{equation}
where $\text{Re}[\Omega^{KS}_{n}]$ is positive defined and $\text{Im}[\Omega^{KS}_{n}] \to 0^-$ to ensure the correct time ordering. The sum is truncated at a finite number of transitions ($N_T$) determined by the number of bands included in the calculation.

The MPA approach provides an analytic continuation for the dressed polarizability $X$ to the complex frequency plane, $z \equiv  \omega + i \varpi$, by representing it as a sum of a few complex poles $n_p$ (usually of the order of 10 to 15), as 
\begin{equation}
    X^{\text{MP}}(z) = \sum_{n}^{n_p} \frac{2 R_n \Omega_n}{z^2-\Omega_n^2}.
    \label{eq:Xmp}
\end{equation}
Note that this representation is applied to each matrix element in reciprocal space, $X^{MP}_{\mathbf{G} \mathbf{G}'} (\mathbf{q},z)$.

By considering Eq.~\eqref{eq:Xmp} and the Lehmann representation for $G_0$, the correlation part of the $GW$ self-energy is then integrated analytically and reads:
\begin{multline}
       \Sigma^{\text{MP}}_c(\omega) = \sum_{m}^{N_B} \sum_{n}^{n_p} P_m v R_n \Bigg[ \frac{f_m}{\omega-E_{m}+\Omega_n -i\eta} + \\
        +\frac{(1-f_m)}{\omega-E_{m}-\Omega_n +i\eta} \Bigg]v.
    \label{eq:Sc}
\end{multline}
where $P_m$ are projectors over KS states, $E_m$ their eigenenergies, and $f_m$ their occupations.
The sum-over-states is truncated at the maximum number of bands, $N_B$.
This expression generalises the PPA solution to the case of a multipole expansion for $X(z)$, and bridges between PPA and an exact full-frequency approach by increasing the number of poles in $X$. More details about this procedure can be found in Ref.~\cite{Leon2021PRB}.

\subsection{MPA sampling for metals}
\label{sec:MPA_sampling_metals}

The poles and residues in Eq.~\eqref{eq:Xmp} are obtained by numerically evaluating $X$ for a number of frequencies equal to twice the number of poles and solving the resulting system of equations (see details in Ref.~\cite{Leon2021PRB}). 
Since the number of poles used in the MPA model, $n_p$, is much smaller than the total number of electron-hole transitions of the target polarizability, $N_T$, the representation, and therefore the efficiency of the method, depends critically on the frequency sampling used in the interpolation.
For semiconductors, the so-called {\it double parallel sampling} proved to be the most robust and accurate with respect to FF calculations, with the fastest convergence with respect to the number of poles. It runs along two parallel lines above the real axis:
\begin{equation}
   {s}^{\text{DP}}= \left\{
    \begin{aligned}
    {\bf z^1} \text{: } z^1_n &= \omega_{n} + i \varpi_1  \\
    {\bf z^2} \text{: } z^2_n &= \omega_{n} + i \varpi_2,
    \end{aligned}
    \right.  \qquad n=1,..,n_p
    \label{eq:s_DP}
\end{equation}
The first of the two branches is closer to the real axis (e.g. with $\varpi_1=0.1$~Ha), except for the first point, set exactly at the origin of coordinates, $z^1_1=0$. The second branch is located further away, typically at $\varpi_2=1$~Ha. 
In a simplified view, $X$ sampled along the first line preserves some of the structure of $X$ in a region close to its poles, while $X$ sampled along the second line is simple enough to be described with a few poles, and accounts for the overall structure of $X$. A more detailed description can be found in Ref.~\cite{Leon2021PRB}.

In order to obtain a numerically stable and effective sampling for metals we found that, at variance with the semiconductor case~\cite{Leon2021PRB}, a small shift of the $z^1_1$ point (in the origin) along the imaginary axis is needed,  
resulting in $z^1_1=i\varpi_1$, where $\varpi_1= 10^{-5}\text{ Ha}$. The shift is done in order to avoid numerical instabilities due to intra-band transitions with energies close to zero. This is similar to the PPA implementation for metals~\cite{Marini2009CPC,Sangalli2019JPCM}, which adopts a $10^{-8}\text{ Ha}$ shift, but in this case along the positive real axis instead of the imaginary axis.

A second difference with respect the strategy used for semiconductors concerns the distribution of the frequency sampling of $X$ along the real axis. For semiconductors~\cite{Leon2021PRB}, the frequency sampling is done in non-uniform grids, in particular, a semi-homogeneous partition in powers of 2 that ranges from 0 to $\omega_m$, called linear partition. Here, we generalize it to any possible exponent $\alpha$:
\begin{equation}
   \{\omega_{n}\}_{\alpha}: \left\{
    \begin{aligned}
        \left(0\right) \text{, } n_p = 1  \\
        \left(0,1\right) \times \omega_m \text{, } n_p = 2  \\
        \left(0,\frac{1}{2},1\right) ^\alpha \times \omega_m \text{, } n_p = 3  \\
        \left( 0,\frac{1}{4},\frac{1}{2},1\right) ^\alpha \times \omega_m \text{, } n_p = 4  \\
        \left( 0,\frac{1}{8},\frac{1}{4},\frac{1}{2},1\right) ^\alpha \times \omega_m \text{, } n_p = 5   \\
        \left( 0,\frac{1}{8},\frac{1}{4},\frac{1}{2},\frac{3}{4},1\right) ^\alpha \times \omega_m \text{, } n_p = 6   \\
        \left( 0,\frac{1}{8},\frac{1}{4},\frac{3}{8},\frac{1}{2},\frac{3}{4},1\right) ^\alpha \times \omega_m \text{, } n_p = 7   \\
        ...
    \end{aligned}
    \right.
    \label{eq:w_grid}
\end{equation}
The distribution described on Ref.~\cite{Leon2021PRB} corresponds to $\alpha=1$. As discussed below, there are cases (see for example the case of copper in Fig.~\ref{fig:Xpol_Cu}), in which $X$ presents a more complex structure at low frequencies and therefore a denser sampling grid in that region is convenient. The distribution corresponding to $\alpha=2$ concentrates more points at low frequencies than the linear case, $\alpha=1$, and permits to increase the accuracy of the $X$ description without changing the frequency range, $\omega_m$, or increase the number of poles used in MPA. In this work, we adopt a quadratic partition, corresponding to $\alpha=2$, for Al and Cu, and a linear one, $\alpha=1$, for Na.

\subsection{Intra-band contributions}
\label{sec:metals_intraband}

Despite the success of the $GW$ approximation, systems with metallic screening present specific methodological challenges, one being the inclusion of intra-band transitions~\cite{Wooten1972book,Marini2001PRB}. Specifically, for partially filled bands, there is a non-vanishing probability that an electron is excited within the same band, i.e.~within  states with quantum numbers $\mathbf{k},n$ and $\mathbf{k}-\mathbf{q},m$, with $n=m$}. Notably, these transitions play an important role, for example, in noble metals~\cite{Marini2002PRL,Krystyna_2020}. Both inter- and intra-band transitions contribute to the irreducible polarizability as defined in Eq.~\eqref{eq:X0}. However, the energy of the pole corresponding to intra-band transitions decreases with $\mathbf{q}$  
until it vanishes in the $\mathbf{q} \to 0$ limit. Despite this behaviour, the contribution to the inverse dielectric function in the case of bulk metals is still finite, due to the divergence of the Coulomb potential, which makes $Y=vX$ not vanishing for $\mathbf{q} \to 0$. For this reason, in the case of metals it is important to properly take this term into account, since it cannot be simply evaluated as in the case of the inter-band contributions.

In principle, it is possible to decrease the weight of the $\mathbf{q}=0$ element, that contains only inter-band terms, by systematically increasing the number of $\bf k$-points in the Brillouin zone (BZ) sampling. However, the contributions from the Fermi surface can dramatically slow down the convergence with respect to the $\mathbf{k}$-space sampling~\cite{Methfessel1989PRB}, resulting in spurious gaps at the Fermi level that vanish very slowly with increasing number of $\mathbf{k}$-points~\cite{Cazzaniga2008PRB}. Several approaches to include the intra-band limit have been proposed. The ones based on explicit Fermi-surface integration~\cite{Maksimov1988JPEMP,Lee1994PRB,Marini2001PRB} are, as explained above, computationally expensive since they require dense $\bf k$-grids. Alternatively, analytical models based on a Taylor expansion of the dielectric function in the small-$\mathbf{q}$ region, avoiding explicit Fermi-surface calculations, are able to remove the spurious gap at the Fermi level with a limited number of $\mathbf{k}$-points~\cite{Cazzaniga2008PRB,Cazzaniga2010PRB,Orhan2019JPCM}. Nevertheless, some of them may depend on {\it ad hoc} external parameters.

A common approach to include the missing intra-band contribution relies on the use of a phenomenological Drude-like term added to the head of the irreducible dielectric matrix in the $\mathbf{q}\to0$ limit, $Y_{\mathbf{G}=\mathbf{G'}=0}(\mathbf{q}=0,\omega)$~\cite{Lee1994PRB}. 
In the long-wavelength limit, $\mathbf{q} \to 0$, the Drude term for the independent particle dielectric function can be written in the form~\cite{Allen1977PRB,book_Palik1985,Smith1986PRB,Maksimov1988JPEMP,Lee1994PRB}
\begin{equation}
    Y_{D}(\omega)=\frac{\omega_D^2}{\omega(\omega+i \gamma)} + O[\mathbf{q}^2],
    \label{eq:drude_term}
\end{equation}
where the Drude frequency, $\omega_D$ (see Table~\ref{table:names}), is an input parameter of the model and the relaxation frequency $\gamma$ is usually a free parameter set typically to $\gamma=0.1$~eV. In principle $\omega_D$  
can be determined fully {\it ab-initio}, resorting to very dense $\mathbf{k}$-point grids~\cite{Lee1994PRB,Marini2002PRL} or to an interpolation of the BZ, for instance with Wannier functions~\cite{Kohn1974PRB,Sporkmann1994PRB,Prandini2019ComputPhysCommun,Prandini2019npjComputMater} or the tetrahedron method~\cite{Methfessel1989PRB,Blochl1994PRB,Lee1994PRB,Friedrichnano2022}. Alternatively, experimental values can also be used when available.

In the next Sections we will discuss the possibility to extrapolate a complex plasmon frequency (see Table~\ref{table:names}) in the  $\mathbf{q}\to {0}$ limit from the frequency structure of $Y(\mathbf{q}, \omega)$ at finite $\mathbf{q}$, which in general is a superposition of intra- and inter-band contributions. 
In a second step, we will use a $f$-sum rule~\cite{book_Palik1985} in the same spirit of Ref.~\cite{Lee1994PRB}, in order to estimate the intra-band contribution to the plasmon frequency. We will also propose a simple and virtually zero-cost method to include an approximate treatment of the missing intra-band limit from first-principles, without the need to resort to any add-on model. 

\section{Results and discussion}
\label{sec:results}

In the following, we present the results for three bulk metallic systems highlighting different issues arising when applying the $GW$ approach to metals. We start by studying the case of two simple metals, Al and Na (see e.g. Refs.~\cite{Levinson1983PRB,Jensen1985PRL,Lyo1988PRL} for a description of their band structures). Next, we focus our attention on Cu, a more challenging system whose electronic structure has been thoroughly studied, both experimentally~\cite{Cu_exp,Cu_exp2} and theoretically~\cite{Marini2002PRL,Cu_Au_life,Liu2016PRB,delBen2019PRB}.
The use of PPA for Cu has been shown to be problematic~\cite{Marini2002PRL} and, for this reason, copper is not only an important test case for the application of MPA and the description of intra-band effects, but also provides a better understanding of the applicability of PPA.

As a starting point for our $GW$ simulations, we use DFT calculations performed at the PBE~\cite{Perdew1996} level using scalar-relativistic optimized norm-conserving Vanderbilt pseudopotentials~\cite{Hamann_2013}, as implemented in the Quantum ESPRESSO package~\cite{QE1, QE2}. The kinetic energy cut-off is set to 100, 70, and 150 Ry  for Al, Na, and Cu, respectively. The $\mathbf{k}$-grids were determined by the convergence requirements of the $GW$ calculations, considering, in particular, the specific treatment of the intra-band limit. When reporting quasi-particle energies, we use $\mathbf{k}$-point grids of $16\times16\times16$ for Al and Na, and $12\times12\times12$ for Cu. Moreover, the $GW$ correction to the Fermi level is linearly interpolated from the corresponding corrections to the closer quasi-particles present in the specific $\mathbf{k}$-mesh. 

The DFT results are in good agreement with previous results obtained with the same method~\cite{Liu2016PRB}, and in reasonable agreement with the results reported for Cu in Ref.~\cite{Marini2002PRL}, performed using LDA~\cite{Marini2001PRB}. In fact, the $GW$ results for Cu have shown to be very sensitive to the choice of the DFT starting point~\cite{Liu2016PRB}, though we will not address this point here.
The $GW$ calculations were done using the \yambo~\cite{Marini2009CPC, Sangalli2019JPCM} code. 
The numerical convergence of the $GW$ results has been checked with care, and the resulting parameters, being system dependent, are detailed in the sections below when discussing the results.

\subsection{MPA for simple metals
\label{sec:MPA_simple}}

\begin{table}

\begin{ruledtabular}
\begin{tabular}{llccc}
  
\\[-8pt]
 &                   & {\bf DFT-PBE} & {\bf GW-PPA} & {\bf GW-MPA}  \\[2pt]
  \hline\\[-8pt] 
 Al &$\Gamma_1$        & -11.12     & -10.79  &-10.94 \\
 &$\Gamma_{25'}$       &  12.71     &  12.30  & 12.48 \\
 &$X_{4'}$            &  -2.93     &  -2.91  & -2.86 \\
 &$W_{3}$             &  -0.85     &  -0.83  & -0.82 \\[2pt]
\hline\\[-8pt]
Na &$\Gamma_1$        & -3.27      & -2.85   & -2.97 \\
   &$\Gamma_{25'}$        & 11.76      & 11.19   &  10.81 \\
\end{tabular}
\end{ruledtabular}
 \caption{Al and Na quasi-particle energies (eV) with respect the Fermi level computed within DFT-PBE, GW-PPA, and GW-MPA using a $16\times16\times16$ $\mathbf{k}$-grid including the $\mathbf{q} \to {0}$ intra-band contribution through the CA method.
 \label{table:QP_Al_Na}}
\end{table}

\begin{figure*}
   \centering
    \includegraphics[width=0.8\textwidth]{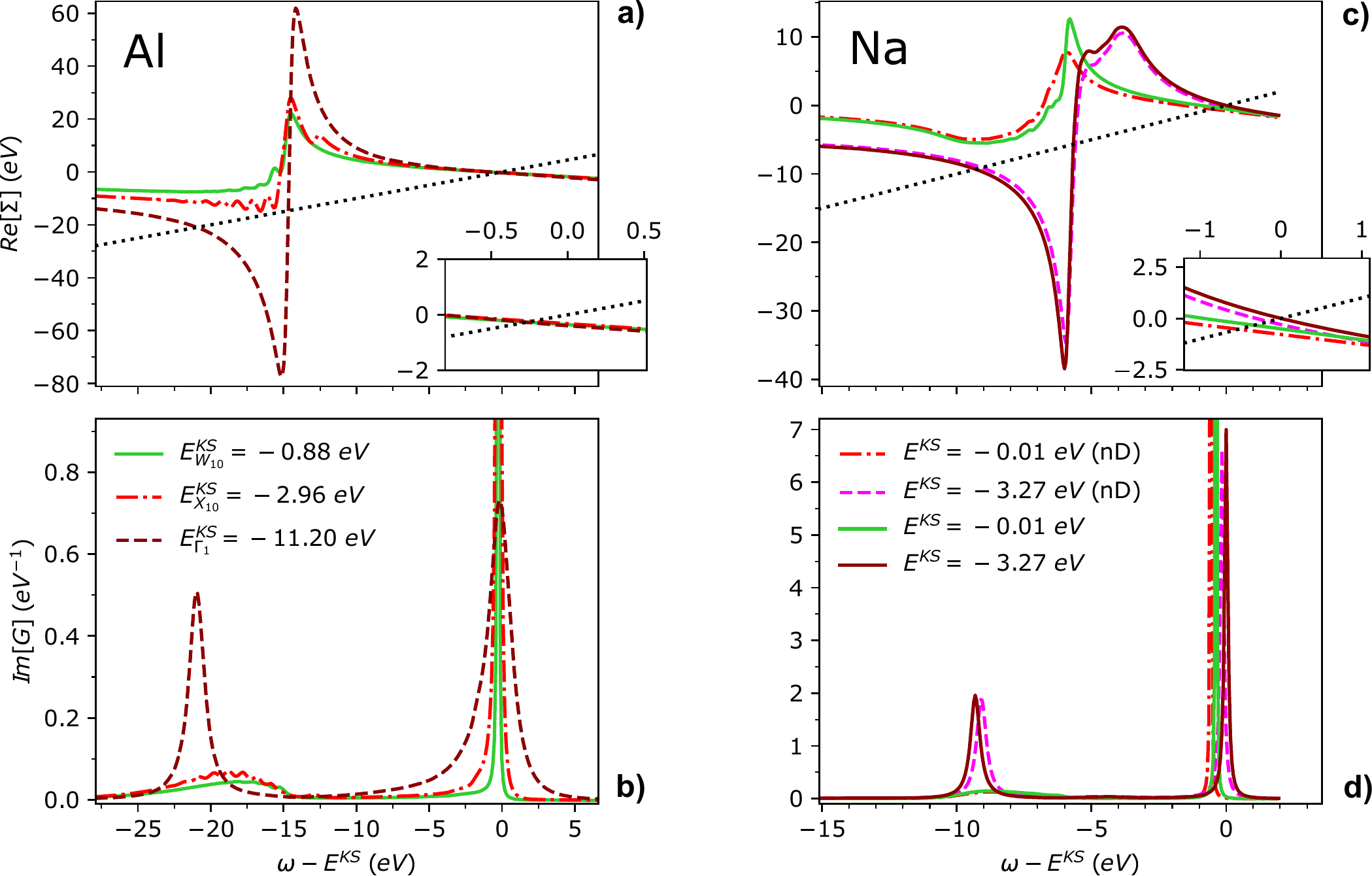}
   \caption{Frequency dependence of the real part of the self-energy (panels a) and c)) and spectral function (panels b) and d)) computed with MPA for three quasi-particles of Al (panels a) and b)) and two of Na (panels c) and d)), including the intra-band limit using CA (see text). In the case of Na, we also show the corresponding curves without any intra-band correction (nD).
   \label{fig:Al-Na_SG}}
\end{figure*}

We start by computing quasi-particle energies of Al and Na using MPA. Here the frequency dependence of the polarizability presents a structure with mainly one strong plasmon peak, similar to that of silicon computed in Ref.~\cite{Leon2021PRB}. As expected, the double parallel sampling ensures convergence with a similar number of poles, $n_p=8$. The present results were obtained considering 300 bands for both $X$ and $\Sigma$ and an energy cut-off for $X$ of 20 and 15~Ry for Al and Na respectively.

In Table~\ref{table:QP_Al_Na} we report the quasi-particle energies for Al and Na, including $\Gamma_1$ (the lowest QP peak at $\Gamma$, corresponding to the valence bandwidth) and other reference quasi-particles, computed using PPA and MPA. MPA QPs are generally in very good agreement with FF values from the literature (see e.g. Ref.~[\onlinecite{Cazzaniga2012PRB}] and references therein). According to our calculations, the computed quasi-particles values for Al and Na with MPA are estimate to differ by less than 8~meV from the corresponding FF-RA results (comparison done using 10 Ry cutoff to represent $X_0$ for both MPA and FF-RA), as found for semiconductors~\cite{Leon2021PRB}. 
Instead, PPA QPs show deviations that are systematically larger for states further from Fermi.

Previous $GW$ calculations for Al and Na~\cite{Cazzaniga2012PRB} have shown that PPA describes well the tail of the self-energy, i.e. the frequency region around the Kohn-Sham energies, and gives reasonable QP solutions for both Al and Na. However, if we consider the whole frequency range, the agreement between PPA and FF-CD is less satisfactory. PPA shows sharp fluctuations in the self-energy and spectral functions, that result in several spurious solutions of the quasi-particle equation, evidenced by multiple small peaks in the spectral function (see e.g.~Fig.~4 of Ref~\cite{Cazzaniga2012PRB}).
In Fig.~\ref{fig:Al-Na_SG} we show the self-energy and spectral function for Al and Na, this time computed with MPA. The comparison with results obtained within FF-CD~\cite{Cazzaniga2012PRB} shows that MPA not only describes well the tail of the $X(\omega)$ and $\Sigma(\omega)$ functions, but also correctly describes the positions of the peaks and their relative intensities in the whole frequency range. 

In the left panels of Fig.~\ref{fig:Al-Na_SG} we focus on Al and plot, as a function of the frequency, the MPA self-energy, $\langle \psi_{m\mathbf{k}} | \Sigma(\omega) | \psi_{m\mathbf{k}}\rangle$, and the spectral function, $\langle \psi_{m\mathbf{k}} | \text{Im}[G(\omega)] | \psi_{m\mathbf{k}}\rangle$. These quantities have been projected on three Al states, one corresponding to the bottom of the valence band at $\Gamma$ and two other Kohn-Sham states closer to the Fermi level. Comparing the three self-energy functions, there is a more effective pole superposition for states at energies further away from the Fermi level. Indeed, for the lowest energy state with E$^{KS}=-11.2$~eV, this leads to a frequency dependence of $\Sigma$ with an intense single pole (at about -15 eV with respect to E$^{KS}$) and consequently a very broad and shallow QP peak in the corresponding spectral function. 
At the same time the satellite structure is enhanced to the point of becoming a second peak, originating from a second solution of the quasi-particle equation (intersections of the dashed line with the self-energy function in the upper panel). This scenario is consistent with the so-called "plasmaron" peak, a sharp satellite feature emerging as an artefact of the $G_0 W_0$ approximation to the self-energy~\cite{martin2016book,Caruso2016EPJB,Guzzo2011PRL}.

The situation is similar for the two QPs computed for Na shown in the rights panel of Fig.~\ref{fig:Al-Na_SG}, with the lowest state presenting again two solutions for the QP equation.

\subsection{Analysis of the intra-band contribution}
\label{sec:intra_analysis}

\begin{figure*}
    \centering
    \includegraphics[width=1\textwidth]{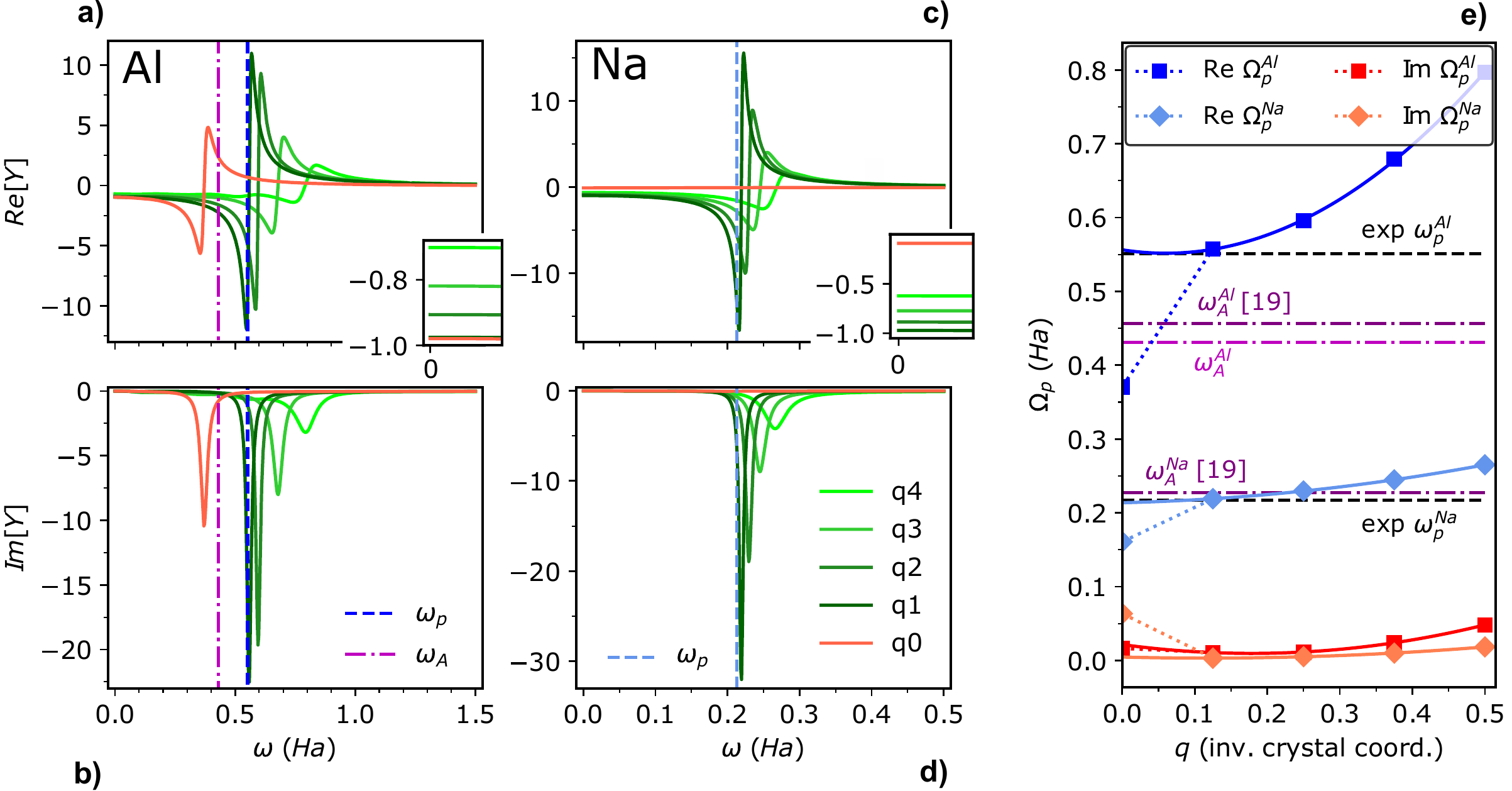}
    \caption{Frequency dependence of $Y_{\mathbf{G}=\mathbf{G'}=0}$ matrix elements computed with MPA for different $\mathbf{q}$ vectors of modulus $q\equiv |\mathbf{q}|$ tending to 0, a) and b) for Al, and c) and d) for Na. For $q=0$ (orange curves) the intra-band transitions are not included. The insets in panels a) and c) show the region around $\omega=0$.
    Panel e) shows the $q$ dispersion of the real and the imaginary parts of the main pole of $Y$ for $q_0$--$q_4$ ($q_n=\frac{n}{8}$ in units of $2\pi/a$, being $a$ the respective Al and Na lattice parameters).
    The solid lines show the corresponding parabolic fits consistent with a Lindhard (bulk) plasmon dispersion~\cite{Lee1994PRB,vomFelde1989PRB,Cazzaniga2008PRB,Huotari2011PRB}. The black dashed lines correspond to the experimental plasmon frequency of Al~\cite{Lee1994PRB} and Na~\cite{vomFelde1989PRB}.
    Dash-dot purple lines correspond to the values of the intra-band frequency, $\omega_A$, computed in Ref.~\cite{Cazzaniga2012PRB} using the method described in Ref.~\cite{Cazzaniga2010PRB}, while the violet one corresponds to our estimate for Al, computed by means of Eq.~\eqref{eq:omega_sum_rule}.
    \label{fig:Xw_Al_Na}}
\end{figure*}

In common $GW$ implementations, especially those targeting semiconductors, the intra-band contribution to the dielectric function in the $\mathbf{q} \to 0$ limit, Eq.~\eqref{eq:X0}, is often not included, as explained in Sec.~\ref{sec:metals_intraband}. 
In the case of Al, where a substantial part of the Fermi surface is very close to the BZ boundary, one can expect~\cite{Cazzaniga2008PRB} that many of the metallic contributions are effectively inter- rather than intra-band terms, resulting in a small error when the intra-band term is neglected~\cite{Cazzaniga2008PRB}, while for Na the intra-band terms are found to be more relevant.

For both Al and Na, in Fig.~\ref{fig:Xw_Al_Na} we show how this affects the frequency dependence of the $Y_{\textbf{G}=\textbf{G'}=0}$ matrix elements computed for different $\mathbf{q}$-vectors along 
an arbitrary direction. The curves in green shades correspond to $Y(\omega)$ computed for finite but small $\mathbf{q}$. The orange curve corresponds to the $\mathbf{q} \to 0$ limit evaluated only for the inter-band term. There are two main differences between the green and orange curves. The first difference is the limit of $\text{Re}[Y]$ as the frequency tends to zero (static limit), that evolves smoothly for finite $\mathbf{q}$ but in general tends to a value different from the one corresponding to $\mathbf{q}=0$. 
 As shown in the insets of Fig.~\ref{fig:Xw_Al_Na}, 
the smallest finite $\mathbf{q}$ provides a static limit very similar to the value for $\mathbf{q}=0$ in the case of Al, while it is considerably larger in the case of Na (both results in agreement with previous studies~\cite{Cazzaniga2008PRB}).

This difference has been commonly used as a measure of the missing intra-band term~\cite{Cazzaniga2008PRB,Cazzaniga2012PRB}, since for metals in the limit $\mathbf{q}\to 0$, $\varepsilon^{-1}_{\mathbf{G}=\mathbf{G'}=0}(\mathbf{q},\omega=0)$ vanishes, meaning that 
$Y_{00}(\mathbf{q},\omega=0) \to -1$,
as apparent from the progression of the curves with finite $\mathbf{q}$, that include intra-band transitions. In fact, in the independent particle picture, the $\mathbf{q} \to 0$ limit of $\text{Re}[Y]$ at $\omega=0$ is related to a non-vanishing probability of vertical transitions within the same band~\cite{Lee1994PRB}, and can therefore be used to estimate a Drude frequency~\cite{Johnson1972PRB,DAmico2020PRB}.
However, this probability alone does not determine the plasmon frequency (see Table~\ref{table:names} for a summary of the nomenclature) or the position of the pole of $\text{Re}[Y]$ for $\mathbf{q} \to 0$.

In fact, the second difference between the orange ({$\mathbf{q}=0$}, no intra-band contribution) and the green curves (finite {$\mathbf{q}$}, intra-band included) in Fig.~\ref{fig:Xw_Al_Na} is the position of the main pole of $Y(\omega)$, here called $\Omega_p$, or in the case of Na, to the apparent absence of poles for {$\mathbf{q}=0$}, whose small amplitudes cannot be seen in the plot.
If the whole frequency range is considered, we see that the behaviour of $\text{Re}[Y(\omega \to 0)]$ depends on the position of $\Omega_p$.
Following the green curves at finite $\mathbf{q}$, it is clear that  $Y_{\mathbf{G}=\mathbf{G'}=0}$ for both Al and Na change smoothly with  $\mathbf{q}$. The curves present a pole, $\Omega_p(\mathbf{q})$, of decreasing energy and increasing amplitude, just above 0.5~Ha for Al and 0.2~Ha for Na. As shown in Fig.~\ref{fig:Xw_Al_Na}(e), both the real and imaginary part of this pole can be easily extrapolated to $\mathbf{q}=0$, by means of the Lindhard plasmon dispersion~\cite{Lee1994PRB,vomFelde1989PRB,Cazzaniga2008PRB,Huotari2011PRB}.

In the same plot we show, as a reference, the Drude frequency corresponding to the {$\mathbf{q} \to 0$} limit of the intra-band contributions, $\omega_A$ (see Table~\ref{table:names}), as computed in Ref.~\cite{Cazzaniga2012PRB} for Al and Na, in addition to the experimental plasmon frequency $\omega_p$ of Al~\cite{Mathewson1972JPFMP,Benbow1975PRB,Krane1978JPFMP,Moller1980PRL,Smith1986PRB,vomFelde1989PRB} and Na~\cite{vomFelde1989PRB}. 
In the simulations we can also extrapolate, already with a $8\times8\times8$ $\mathbf{k}$-point mesh, 
the plasmon frequency {at $\mathbf{q}\to 0$} from the position of the main structure of the response functions, namely $\omega_p \equiv \text{Re}[\Omega_p]$. This procedure provides $\omega_p=0.55$~Ha (15.01~eV) for Al, in excellent agreement with the experimental value of 15.0~eV~\cite{Lee1994PRB}. Similarly, the value extrapolated for Na, $\omega_p=0.21$~Ha (5.79~eV), matches very well the experimental value of 5.9~eV~\cite{vomFelde1989PRB} and both compare well with the Drude intra-band frequency computed in Ref.~\cite{Cazzaniga2012PRB} (6.18 eV). The small difference between our theoretical result and Ref.~\cite{Cazzaniga2012PRB} can be attributed to methodological differences (e.g., the DFT functional on top of which the $GW$ calculations are performed). 
In contrast, the difference between  $\omega_p$ and $\omega_A$ for Al is larger than 2.5~eV since the plasmon frequency $\omega_p$ has non-negligible contributions from both intra- and inter-band transitions, as previously reported in Refs.~\cite{Smith1986PRB,Lee1994PRB}. Note however that the inter-band contributions are not included in the Drude frequency computed in Ref.~\cite{Cazzaniga2012PRB}.

In order to discriminate between the intra- and inter-band contributions to the plasmon frequency, we have used a simple expression based on the $f$-sum rule~\cite{Stefanucci-vanLeeuwen2013book,martin2016book,Lee1994PRB}, but separating the two contributions:
\begin{equation}
    \Omega_A^2 = \lim_{\mathbf{q} \to {0}} \frac{2}{\pi} \int_{0}^{\infty} d\omega \, \omega \, \text{Im}[Y(\mathbf{q},\omega)-Y_E(\mathbf{q},\omega)],
    \label{eq:freq_sum_rule}
\end{equation}
where $Y_E$ corresponds to inter-band transitions only, while $Y$ accounts for the complete response. Within MPA the integral is solved analytically (derivation in Sec.~I of the \suppinfo), leading to:
\begin{equation}
    \Omega_A^2 = 2v(R_p\Omega_p -R_E\Omega_E),
    \label{eq:omega_sum_rule}
\end{equation}
where $\Omega_E$ and $v R_E$ are the position and the residue of the most relevant pole of $Y_E(\mathbf{q}=0)$, while $\Omega_p$ and $v R_p$ are the corresponding values for $Y(\mathbf{q}=0)$. 

In principle, the product $v R_p \Omega_p$ should be computed in the $\mathbf{q} \to 0$ limit. We have instead considered the extrapolation of $\Omega_p^2$, which is equivalent in our model (see Sec.~I in the \suppinfo) and significantly more stable.  
The values of $vR_E\Omega_E$ are taken directly from the calculation at $\mathbf{q}=0$ (orange curves in Fig.~\ref{fig:Xw_Al_Na}a,b), since no intra-band transitions are considered, as explained above.
For Al, the real part of $\Omega_E$  is $\omega_E=0.37$~Ha (10.08~eV) and thus, applying Eq.~\eqref{eq:omega_sum_rule}, the real part of the intra-band pole $\Omega_A$ is $\omega_A=0.43$~Ha (11.72~eV). For Na, $\omega_p$ and $\omega_A$ are similar.
The comparison of $\omega_p$ and $\omega_A$ confirms that the experimental plasmon frequency, $\omega_p$, in the case of Na corresponds mainly to intra-band contributions, while for Al there is an important inter-band contribution~\cite{Smith1986PRB}, and its use as a Drude intra-band frequency would result in an overestimation of the actual $\omega_A$.

\begin{table}
\begin{ruledtabular}
\begin{tabular}{lcc} 
\\[-8pt]
 {\bf Contribution}        & {\bf Pole (complex)} & {\bf Frequency (real)} \\
  \hline\\[-8pt] 
 intra-band      & $\Omega_A$& $\omega_A = Re[\Omega_A]$ \\
 inter-band       & $\Omega_E$ & $\omega_E= Re[\Omega_E]$ \\
 plasmon(intra+inter)          & $\Omega_p$ & $\omega_p = Re[\Omega_p]$\\
 plasma           & -          & $\omega_\text{pl}=\sqrt{4\pi \rho_e}$ \\
 Drude(model)    & $\omega_D + i \gamma$ & $\omega_D$ \\
\end{tabular}
\end{ruledtabular}

 \caption{Summary of the notation concerning frequency related quantities introduced in this work. The plasma frequency is defined in terms of the electronic density, $\rho_e$. The Drude pole/frequency are model parameters used to describe the plasmon or only its intra-band contribution, as described in Eq.~\eqref{eq:drude_term}. 
 \label{table:names}}
\end{table}

Making use of the extrapolation procedures described above in the context of the MPA framework, and of a simple $f$-sum rule, it is possible to determine not only the real but also the imaginary part of both the plasmon and the intra-band pole, usually not considered in other {\it ab-initio} methods. 
It is also worth noticing that the extrapolation is done with points from a much coarser $\mathbf{k}$-grid ($8\times8\times8$ for both Al and Na), with respect to the grids required to compute the intra-band frequency with an independent particle formulation~\cite{Marini2001PRB, Cazzaniga2008PRB}.

Despite the limited accuracy of the computed imaginary values, they are meaningful and provide a qualitative understanding of how intra- and inter-band terms, linearly summed at the independent particle level, are combined after the inversion of the Dyson equation. While the Na case is trivial, since the inter-band contribution is negligible, in the case of Al the small difference between $\omega_A$ and $\omega_E$, comparable to their imaginary parts, explains the presence of a single pole in $Y(\omega)$ located roughly at $\omega_p^2 \sim \omega_A^2 +\omega_E^2$ (see Sec.~I of \suppinfo).

\subsection{Modelling of the intra-band limit}
\label{sec:intra_modelling}

Our analysis of the dressed response function $Y(\omega)$ suggests that an alternative to the direct evaluation of the intra-band limit, usually determined from $X$ at the independent particle level~\cite{Marini2001PRB},
can be obtained, either by 
(1) including a complex Drude pole $Y_D(\omega)$, according to Eq.~\eqref{eq:drude_term}, in the head ($\mathbf{G}=\mathbf{G'}=0$) of the independent particle dielectric function, with the Drude frequency given by the computed intra-band pole; or
(2) approximating the full $Y(\mathbf{q}=0)$ matrix element by its nearest neighbour $Y(\mathbf{q}\neq0$), i.e. with the $\mathbf{q}$-vector closest to 0 according to the adopted $\mathbf{k}$-point grid. 

The first method builds on using an estimate of the Drude intra-band frequency, similar to the extrapolations used in~\cite{DAmico2020PRB}, but here considering the whole frequency range and both intra- and inter-band contributions. The second method, which we will call from now on {\it constant dielectric function approximation} (CA), assumes that the whole $Y(\mathbf{q})$ matrix is constant in a small region around $\mathbf{q}=0$. 
This approach is inspired by the leading term of the Taylor expansion for small $\mathbf{q}$ of the Thomas-Fermi distribution, and is corroborated by the small difference of 0.006~Ha (0.17~eV) found for both, Al and Na, between the extrapolated value of $\Omega_p$ and its value at the first finite $\mathbf{q}$, as shown in  Fig.~\ref{fig:Xw_Al_Na}~e). Both methods simultaneously correct the position of the plasmon pole and the limit of $Y_{\mathbf{G}=\mathbf{G}'=0}$ for $\omega=0$ and add virtually no computational cost to the calculation. In addition, CA also corrects other matrix elements for which the intra-band limit may be important.

In Sec.~II of~\suppinfo{} we report plots similar to the ones in Fig.~\ref{fig:Xw_Al_Na} for $Y$ matrix elements of Na other than the head, showing that after the head ($Y_{\mathbf{G}=\mathbf{G'}=0}$), intra-band contributions are relevant also for the so-called wing elements ($Y_{\mathbf{G}=0\neq \mathbf{G}'}$ and $Y_{\mathbf{G}\neq 0 =\mathbf{G}'}$), while less important for the diagonal elements ($Y_{\mathbf{G}=\mathbf{G}'\neq 0}$), specially at increasing $|\mathbf{G}|$. For finite $|\mathbf{G}|$ the evolution of the $Y(\mathbf{q})$ matrix elements when $ \mathbf{q}\to 0$ is less smooth and the position of the poles does not always change monotonously, meaning that an extrapolation would require a denser $\mathbf{k}$-point grid. Even if the constant dielectric function approximation has limited accuracy for some of these matrix elements, it still provides a significant overall improvement. In particular in materials such as Cu, as discussed below, the CA method presents some clear advantages regarding the estimation of $\omega_A$.

\begin{figure}
    \centering
    \includegraphics[width=0.46\textwidth]{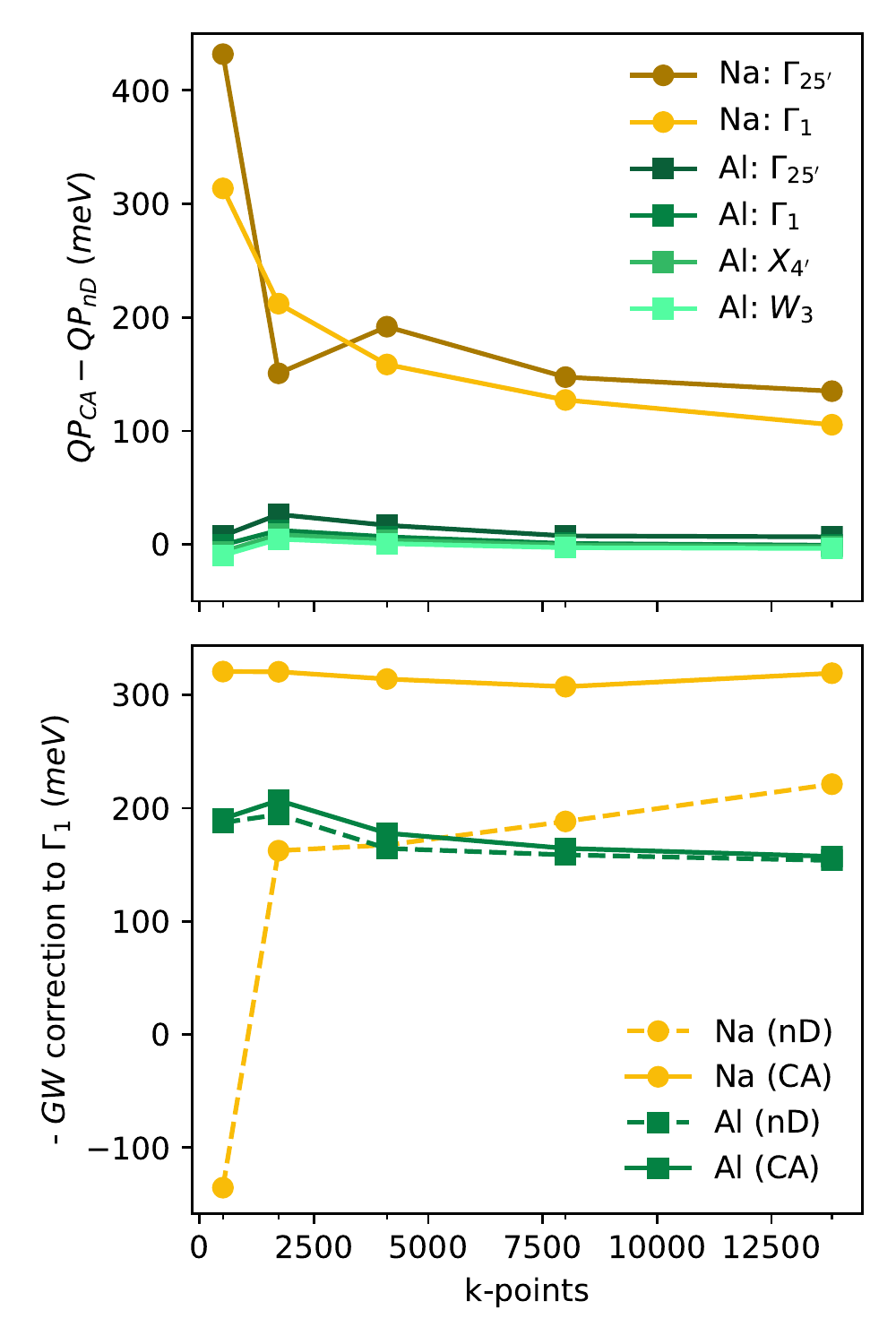}\\
    \caption{Top panel: Difference between GW-MPA corrections computed with the (CA) intra-band term and without (nD) as a function of the number of $\mathbf{k}$-points, for 2 quasi-particles of Na (light and dark yellow) and 4 of Al (green shades). Bottom panel: Convergence of the $GW$ correction for the QP at $\Gamma_1$ of Na (yellow) and Al (green) with CA (solid) and nD (dashed).}
    \label{fig:Drude_Al-Na}
\end{figure}

To assess the effect of this approximation in the QP solution, in Fig.~\ref{fig:Drude_Al-Na} we show Al and Na QP energies computed without (nD) and with (CA) intra-band corrections.
When the number of $\mathbf{k}$-points is increased, the weight of the $Y(\mathbf{q}=0)$ element in the self-energy decreases and both methods eventually converge to the same quasi-particle values, but only very slowly, as discussed above. In fact, Fig.~\ref{fig:Drude_Al-Na} shows that for two selected QPs of Na the intra-band term is fundamental due to the importance of this contribution to the screening properties of the system. In contrast, for Al the difference is small, and the convergence is governed by the inter- rather than intra-band contributions for all the 4  QPs considered. 
In the bottom panel of Fig.~\ref{fig:Drude_Al-Na} one can see the significant acceleration introduced by CA in the convergence of the bandwidth of Na, where, besides a small oscillation in the $20\times20\times20$ grid (caused by oscillations in the DFT eigenvalues), the first point corresponding to the $8\times8\times8$ mesh already provides very accurate results.
In the CA scheme the convergence benefits simultaneously from the decrease of the weight of the $Y(\mathbf{q}=0)$ contribution and from the fact that the correction itself improves for denser grids in reciprocal space, since the first $\mathbf{q}\neq0$ is closer to 0.

In Fig.~\ref{fig:Al-Na_SG} we show the frequency dependence of the real part of the self-energy (top) and spectral function (bottom) computed for two quasi-particles of Na, within MPA with and without the intra-band correction. The correction does not change dramatically the shape of the self-energy, but introduces an extra pole in the real part of the self-energy at the intra-band frequency ($\sim$-6~eV) 
and renormalizes the peaks of the spectral function. 
The inclusion of this term promotes the pole overlapping around the plasmon frequency, affecting the tail of the self-energy and thus the QP solution as illustrated in the insets of Fig.~\ref{fig:Al-Na_SG}, differently for each quasi-particle.  

In the case of Al and Na, the QP energies computed with the Drude model, Eq.~\eqref{eq:drude_term} using as input $\omega_D=\omega_A$, and the CA schemes are very similar, with differences below 20~meV when using the $8\times 8\times 8$ $\mathbf{k}$-grid. This leads us to conclude that the CA scheme could replace the usual Drude correction, replacing a semiempirical scheme by a simple {\it ab-initio} approximation. This is particularly relevant when the Drude intra-band frequency is difficult to estimate either from experiments or calculations, since the CA scheme has virtually zero computational cost and, as the extrapolation presented in the previous Section, describes both the real and the imaginary part of $Y$.

To summarize this section, the inclusion of the intra-band limit through the proposed CA scheme requires no extra computational cost with respect to the standard $GW$ calculation and accelerates the $\mathbf{k}$-grid convergence of the QP energies for systems where the intra-band contribution dominates, like Na, without resorting to semi-empirical corrections such as the Drude model or computationally costly {\it ab-initio} approaches.

\subsection{Frequency representation of the response function of copper}
\label{sec:Cu_ppa}

\begin{figure*}
    \centering
    \includegraphics[width=0.99\textwidth]{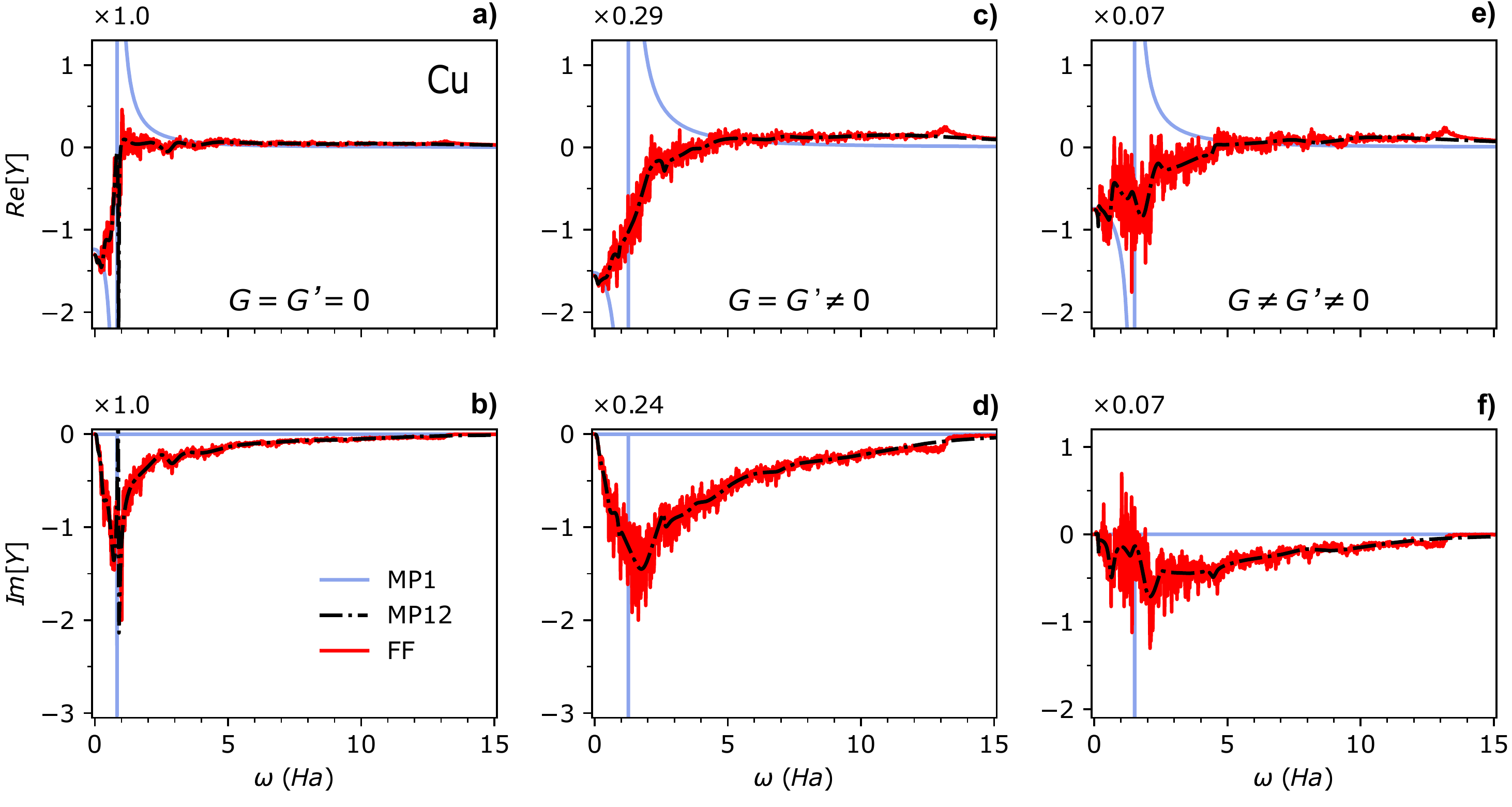}
    \caption{Selected Cu $Y(\mathbf{q}=0)$ matrix elements computed within MPA with 1 and 12 poles compared with the corresponding FF results. The y-axes are scaled with the factors indicated on top of each panel.}
    \label{fig:Xpol_Cu}
\end{figure*}

As mentioned before, the case of copper presents several challenges for an accurate $GW$ description. 
The Cu band structure features a series of flat $d$-bands around 2~eV below the Fermi level, leading to strong transitions in $Y_{\mathbf{G},\mathbf{G}'}(\mathbf{q},\omega)$ spread over a large energy range~\cite{Marini2002PRL}. 
As shown in Fig.~\ref{fig:Xpol_Cu} for $\mathbf{q}=0$, even for small values of $\mathbf{G}$ and $\mathbf{G}'$, $Y_{\mathbf{G},\mathbf{G}'}(\mathbf{q},\omega)$ can behave very differently from a single pole case, hindering the use of PPA but suggesting that a multipole approach could prevent resorting to more expensive FF methods.

When considering PPA or in general MPA with only a few poles, one of the main issues is that the interpolation of $X$ or $Y$ may give rise to non-physical poles, posing representability problems. 
Within the Godby and Needs (GN) PPA scheme implemented in \yambo~\cite{Marini2009CPC,Rangel2020CPC,Godby1989PRL,Oschlies1995PRB}, 
the condition used to identify these so-called unfulfilled modes is the following:
\begin{equation}
    \text{Re}\left[ \frac{ Y_{\mathbf{G}\mathbf{G}'}(\mathbf{q},0)}{Y_{\mathbf{G}\mathbf{G}'}(\mathbf{q},i \varpi_\text{pl})} -1 \right] <0, 
    \label{eq:PPcond}
\end{equation}
$\varpi_\text{pl}$ being a frequency on the imaginary axis used to perform the GN interpolation, typically set to $\varpi_\text{pl}=1$~Ha or to a value of the order of the plasma frequency ($\varpi_\text{pl} \gtrsim \omega_\text{pl}$), computed from the electronic density, $\rho_e$ (see Table~\ref{table:names}).
As an example, for the diagonal elements ($\mathbf{G}=\mathbf{G}'$), the polarizability evaluated on the imaginary axis should be real and therefore unfulfilled modes are those for which the resulting pole is instead imaginary. In these cases, the position of the pole is typically set to $\Omega_{\text{fail}}^{\text{GN}}=1$~Ha.

Setting the pole at $\Omega_{\text{fail}}^{\text{GN}}$ usually works well for simple semiconductors~\cite{Rangel2020CPC,Leon2021PRB}. However, in more complex systems it can compromise the PPA approach. In fact, when performing $GW$ calculations using GN-PPA for Cu, we found that no less than $48\%$ of the matrix elements are unfulfilled modes.
This means that, for almost half of the matrix elements, the position of the pole is spuriously set to 1~Ha, severely affecting the self-energy and the quasi-particle solution, as shown in the insets of Fig.~\ref{fig:Sg_Cu}.
Within MPA, increasing the number of poles in the description of $Y$, together with the generalized condition to assign the position of the poles of the unfulfilled modes, as described in Ref.~\cite{Leon2021PRB}, leads to a significant improvement in the representability of $Y$, as illustrated in Sec.~III of \suppinfo.

In Fig.~\ref{fig:Xpol_Cu} we compare selected $Y$ matrix elements computed within MPA with 1 and 12 poles, with the FF results computed with a frequency grid of 1000 points (all other convergence parameters being the same: $\mathbf{k}$-grid, number of empty bands, etc)
At first glance, the enveloping structure of diagonal elements presents a strong overall peak, as in the case of  semiconductors such as Si, hBN, and TiO$_2$, which are well-described   within PPA and MPA~\cite{Leon2021PRB}. However, in the case of Cu, there are other important peaks close to the origin  not captured by a single-pole model. In this case, PPA quasi-particle energies are not just numerically inaccurate, as in the case of the discussed semiconductors, but PPA becomes an inadequate model. Increasing the number of poles from 1 to 12 significantly improves the agreement between $Y$ computed with MPA and FF, reproducing the overall frequency dependence even if MPA presents a much smoother shape. 

While the rapid oscillations in the FF response function are enhanced by the discretization of the Brillouin zone,
the origin of such fluctuations can be related to the topology of the flat $d$-bands of Cu~\cite{Marini2002PRL}, 
consistently e.g. with the very structured $W(\omega)$ computed for Ni~\cite{Ni_X_1998}. In fact, regardless of the overall simple shape of $X$, numerous inter-band transitions, close in energy and not effectively overlapped, contribute to the fluctuations of the polarizability $X$ and of the inverse dielectric function $Y$, when computed within FF. Nevertheless, as discussed in the next Section, they do not significantly influence the computed $GW$ quasi-particle energies.

\subsection{Quasi-particles and spectral function of copper
\label{Sec:Sigma_Cu}}

 \begin{figure*}
    \centering
    \includegraphics[width=0.999\textwidth]{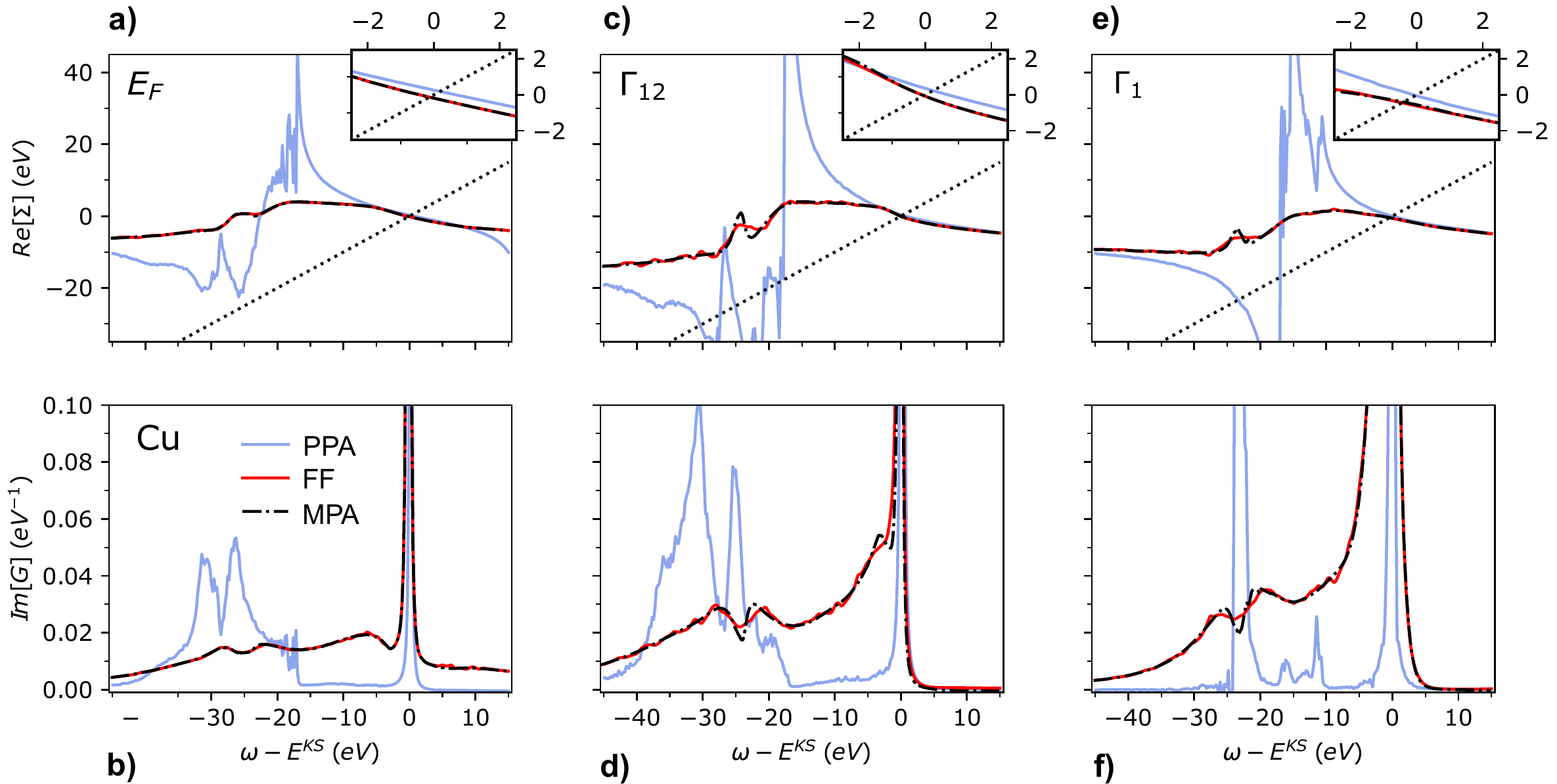}
    \caption{Frequency dependence of the real part of the self-energy (top) and spectral function (bottom) of three quasi-particle states of Cu: one close to the Fermi energy (panels a) and b)), $\Gamma_{12}$ (c) and d)) and $\Gamma_{1}$ (e) and f)); computed with PPA, MPA and FF.}
    \label{fig:Sg_Cu}
\end{figure*}

\begin{table*}
\center
\begin{ruledtabular}
\begin{tabular}{lp{5pt}ccccccccc}
  \\[-5pt]
   QP(eV)                && DFT/LDA & DFT/PBE & DFT/PBE      & GW@LDA & GW@PBE  & GW@PBE    & Exp        \\
                     && Ref.~\cite{Marini2002PRL} & Ref.~\cite{Liu2016PRB} & (current work)      & Ref.~\cite{Marini2002PRL} & Ref.~\cite{Liu2016PRB}  & (current work)    & Ref.~\cite{Cu_exp}        \\[5pt]
  \hline\\[-3pt] 
   $\Gamma_{12}$         && -2.27                & -2.05    &-2.18          & -2.81            & -1.92 to -2.11                & -2.12         & -2.78                       \\ 
   $\Gamma_{1}$          && -9.79                & -9.29    &-9.27          & -9.24            & -9.14 to -9.20                & -9.06         & -8.60                       \\ 
   $X_5$                 && -1.40                & -1.33    &-1.49          & -2.04            & -1.45 to -1.22                & -1.39         & -2.01                       \\ 
   $L_{2'}$              && -1.12                & -0.92    &-0.99         & -0.57             & -0.98 to -1.02                & -1.05         & -0.85                       \\
   $L_3$                 && -1.63                & -1.47    &-1.63          & -2.24            & -1.58 to -1.36                & -1.57         & -2.25                       \\  
   $L$ gap               && 5.40                 & 4.80     &4.66          & 4.76              & 4.98 to 5.09                  &  4.88         & 4.95                        \\[5pt]
 \end{tabular}
 \end{ruledtabular}
 \caption{DFT and $GW$ quasi-particle energies of Cu computed with different methodologies by different groups and compared with the experimental values. All the $GW$ calculations correspond to FF approaches ran on top of LDA~\cite{Marini2002PRL} and PBE~\cite{Liu2016PRB}.} 
  \label{tab:Cu}
 \end{table*}

In Fig.~\ref{fig:Sg_Cu} (top panels) we show the frequency dependence of the self-energy projected on three  selected quasi-particle states of Cu calculated within PPA, MPA, and FF-RA. 
The details of $\Sigma$ computed within the FF approach, better appreciated in Fig.~\ref{fig:Sg_Cu}~c), depend on the fine structure of $W$, which requires a dense frequency grid when computing the polarizability, as shown in Sec.~V of \suppinfo.
Since these calculations are very expensive, the curves shown in Fig.~\ref{fig:Sg_Cu} were computed including 200 bands for all the three methods, and using a frequency grid with 1000 points for FF and no intra-band correction. 
Fully converged MPA results and intra-band corrections are discussed at the end of this Section.

The FF self-energy presents a rather flat structure with no dominant peaks.
Since $\Sigma$ is obtained from the convolution of $G$ and $W$ in Eq.~\eqref{eq:GW}, the oscillations of $W$ are attenuated, resulting in a much smoother function. Nevertheless, the convergence of the QP solution is challenging, since it requires an accurate description of the tail of the self-energy, as shown in the insets of Fig.~\ref{fig:Sg_Cu}. 
This could explain, at least in part, the variety of results present in the literature.

GW-PPA data (blue curves in Fig.~\ref{fig:Sg_Cu}) show that the quasi-particle solution (insets of Fig.~\ref{fig:Sg_Cu}) obtained with a single pole model for $W$ deviates from the FF solution. Besides the deviations at the tail of $\Sigma$, PPA fails to describe the frequency dependence of $\Sigma$ and the spectral function (bottom panels). On the other hand, the MPA results, here obtained with 12 poles and the quadratic sampling, are very accurate, not only in the tail region, that determines the QP corrections, but also for the whole frequency range of both the self-energy and the spectral function.

Comparing the three selected quasi-particle states in Fig.~\ref{fig:Sg_Cu}, the effect of the overlapping of the independent-particle excitations (due to the inclusion of local field effects via the Dyson equation for $W$) on the self-energy of Cu is more relevant for $\Gamma_{1}$ than for $\Gamma_{12}$ and the QPs around the Fermi energy. 
Indeed, as shown in the bottom panels of Fig.~\ref{fig:Sg_Cu}, for the QPs closer to the Fermi level, the shape of the spectral function has a very narrow quasi-particle peak and three satellite. When compared to the QPs close to Fermi, the QPs at deeper energies ($\Gamma_{12}$ and $\Gamma_{1}$) present a broader quasi-particle peak and more intense satellites. The shallower satellite (above -10~eV) forms a shoulder structure for $\Gamma_{12}$(central panel) and eventually merges with the QP peak to form a single broader peak for $\Gamma_{1}$ (right panel).
Despite its complexity, the Cu states at different energies present similar trends as the cases of Al and Na discussed in Sec.~\ref{sec:MPA_simple}.

It is worth to emphasize the importance of the frequency sampling in MPA.
Since copper $X$ and $Y$ present a rich structure at low frequencies, but the energy range, $\omega_m$ in Eq.~\eqref{eq:w_grid} is still large, the quadratic sampling has shown to be more efficient than the linear one. Specifically, it provides, with the same number of poles and the same $\omega_m$, a larger density of points in the low frequency region and therefore higher accuracy. The comparison between the computational cost of MPA and the FF-RA method can be done in a simplified way by comparing the number of frequencies for which $X$ is numerically computed in each approach. Here, for MPA we use 24 frequency points, corresponding to 12 poles, while the FF-RA frequency grid has 1000 points, corresponding to a 40 times gain in computational efficiency of MPA with respect to FF-RA.

The convergence with respect to the number of bands and the size of the $X$ matrices is particularly challenging, as already reported for example for other systems with {\it d} states~\cite{Shih2010PRL,Friedrich2011PRB,Berger2012PRB}, with a slow, non-monotone convergence that hinders the use of extrapolations (more detail in Sec.~V of \suppinfo). For this reason, the computational efficiency of MPA is particularly beneficial as it allows for the use of fine $GW$ convergence parameters, thereby increasing the overall accuracy of the results.

In Table~\ref{tab:Cu} we show the MPA results obtained with 60 Ry of energy cut-off and 1000 bands for both, $X$ and $\Sigma$. These parameters are comparable to the largest ones used within a static subspace approximation~\cite{delBen2019PRB}.
The reported MPA quasi-particle energies are in good agreement with previous calculations using different FF approaches, and summarized in Table~\ref{tab:Cu}. The main differences can be explained by the use of different starting points for the $GW$ calculation, i.e. different exchange-correlation functionals and/or pseudopotentials in the DFT ground state, and possibly to an incomplete convergence of some of the results. While the use of converged parameters is essential when comparing the computed QP energies with experiments, $GW$ corrections do not always improve over DFT/PBE results, as also observed in Refs.~\cite{Liu2016PRB,delBen2019PRB}. In the present case, $GW$  significantly improves $\Gamma_{1}$, while for $\Gamma_{12}$ and other QPs, the $GW$ correction is rather small and slightly worsens the DFT results. The localized nature of the $d$ states in Cu may require methods beyond $GW$ in order to further improve the agreement with  experiments~\cite{Marini2002PRB,Rangel2012PRB,Zhou2020PNAS}.


\subsection{Intra-bands effects in copper}
\label{Sec:Intra_Cu}

\begin{figure*}
    \centering
    \includegraphics[width=0.80\textwidth]{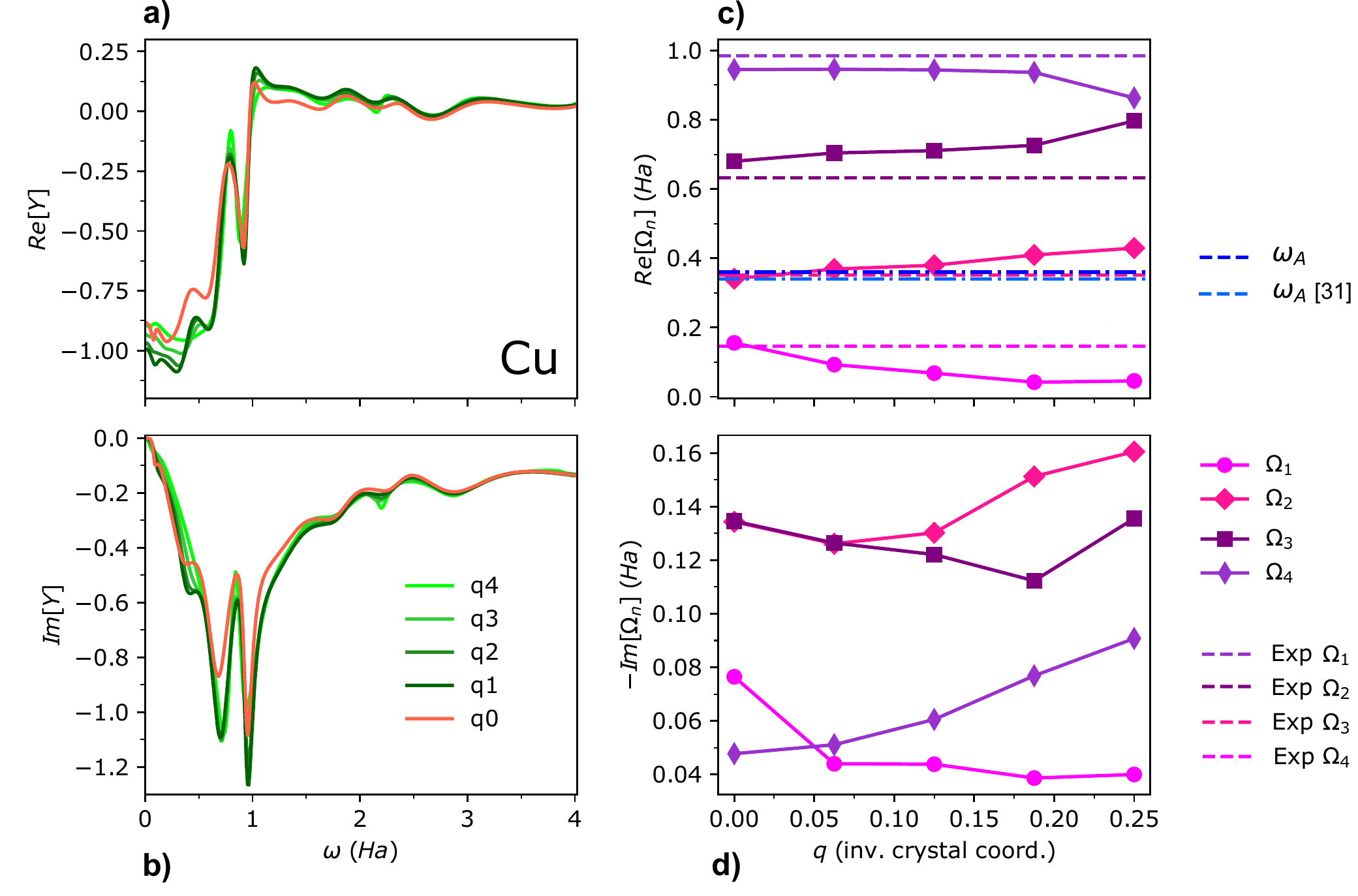}
    \caption{Left panels: frequency dependence of the real (a)) and imaginary part (b)) of $Y_{\mathbf{G}=\mathbf{G}'=0}$ for Cu computed within MPA for different $q$-values tending to 0 ($q_n=\frac{n}{16}$ in units of $2\pi/a$, where $a$ is the lattice parameter of Cu). For $q=0$ (orange curves) the intra-band term is not included. Right panels: real (c)) and imaginary part (d)) of the four most relevant poles at low energies in the $Y$ curves for different $q$ values. The purple dashed lines lines correspond to the position of the poles extracted from optical measurements collected in Ref.~\cite{book_Palik1985}, as explained in the main text. The blue dashed lines correspond to the values of the intra-band frequency, $\omega_A$, computed by means of Eq.~\eqref{eq:omega_sum_rule}, and reported in Ref.~\cite{Marini2001PRB}.
    }
    \label{fig:drude_Cu}
\end{figure*}

In order to investigate the intra-band contributions of copper, in Fig.~\ref{fig:drude_Cu} we show the frequency dependence of the $Y_{\mathbf{G}=\mathbf{G}'=0}$ matrix elements computed for for the smallest $\mathbf{q}$-vectors along one direction of a $16\times 16\times 16$ $\mathbf{k}$-grid. Since $Y(\omega)$ of Cu is very structured at small frequencies, where the effects of the intra-band contributions are expected to be stronger, we have used MPA with a quadratic sampling, Eq.~\eqref{eq:w_grid} with $\alpha=2$ and $n_p=15$, a number of poles slightly larger than the value needed to converge the quasi-particle energies. In contrast with Na, the orange curve ($\mathbf{q}=0$, no intra-band contribution) presents a similar shape and scale with respect to the green curves (small but finite $\mathbf{q}$, with intra-band contributions), even if with less intense peaks. 

In the right panel of Fig.~\ref{fig:drude_Cu} we show the position of the first 4 poles of $Y(\omega)$ as a function of $\mathbf{q}$, which present a rather flat dispersion, when compared with the plasmon dispersion of Al in Fig.~\ref{fig:Xw_Al_Na}. As expected, for $\mathbf{q}=0$ the position of some poles does not correspond exactly to the limit given by the curves with finite $\mathbf{q}$. However, the main difference between the zero and finite $\mathbf{q}$ curves of $Y(\omega)$ is not in the position but rather in the value of the residues of the poles, which is reflected in the intensity of some of the peaks, as shown in Fig.~\ref{fig:drude_Cu}.

In order to compare the computed results with experiments, we used electron energy loss data extracted from a compilation of optical measurements found in Table 1 of the Chapter {\it Optical constants of metals} of Ref.~\cite{book_Palik1985} (see e.g. Fig. 8 of Ref.~\cite{Marini2001PRB}), after interpolation with a multipole model. For this, we chose 18 points of the spectra, with a frequency distribution corresponding to the quadratic sampling of Eq.~\eqref{eq:w_grid} and used them to interpolate a 9 pole model. We then analysed the 4 poles with the highest residues in the frequency interval we are interested in.
In the upper panel of Fig.~\ref{fig:drude_Cu} we show, as horizontal lines, the corresponding experimental energies of the poles. Interestingly, the experimental poles are very similar to the poles computed at the RPA level within MPA. This supports the interpretation that the MPA poles of $Y$ are not a mere mathematical construct aimed at improving representability but indeed correspond to physical collective excitations, each of them describing the envelope of a set of single particle transitions, with a finite imaginary part corresponding to the width of the excitation.
We emphasize that the agreement with the experiment is achieved without resorting to any {\it ad hoc} parameters such as the damping in the case of the FF-RA method of Ref.~\cite{Marini2001PRB}.

In simpler systems, the inclusion of the intra-band limit, even with a simple Drude tail fitted from the experimental spectra, is expected to correct the residues and thus the intensity of the peaks at $\mathbf{q}=0$. 
However, in systems for which the intra-  and inter-band contributions are superimposed in a more structured frequency dependence, the description of the experimental spectra with only the Drude term from Eq.~\eqref{eq:drude_term} 
is not possible~\cite{Suffczynski1960PR,Johnson1972PRB,Allen1977PRB}, 
and indeed models often resort to variable or multiple relaxation frequencies~\cite{Suffczynski1960PR,Benbow1975PRB,Allen1977PRB}. 
In fact, as shown in Fig~\ref{fig:drude_Cu}, the $\mathbf{q}$ dependence of $Y$ does not allow one to 
discriminate between peaks with an intra- or an inter-band character.
In order to circumvent this difficulty, in Ref.~\cite{Marini2001PRB} the intra-band frequency is evaluated numerically as the limit of an intra-band integral at the independent particle level, while in Ref.~\cite{Orhan2019JPCM} it is estimated within a non-interacting uniform-gas theory. 

Here we use again the $f$-sum rule by integrating Eq.~\eqref{eq:freq_sum_rule}, but generalizing Eq.~\eqref{eq:omega_sum_rule} to the case where, in contrast with Al and Na, more than one pole contributes to the intra-band term (see Sec.~I of \suppinfo).
The resulting intra-band frequency, 
$\omega_A=0.36$~Ha (9.80~eV), compares well with the corresponding result of $0.34$~Ha (9.27~eV) from Ref.~\cite{Marini2001PRB} and both values are very close in energy to the second pole shown in Fig.~\ref{fig:drude_Cu}.
We find that intra-band contributions represent around the $25\%$ of the corresponding $f$-sum rule of this pole ($R \Omega$ product), being the largest ratio among all the poles. However, as can be appreciated in Fig.~\ref{fig:drude_Cu} from the change of intensity of the peaks, the inter-band contributions are dominant. In fact, the intra-band contributions to the total $f$-sum rule (sum of all $R \Omega$ products) is rather small, less than $4\%$. 

Using the frequency determined in Ref.~\cite{Marini2001PRB} (9.27~eV) and the relaxation frequency fixed to 0.1~eV as the inputs to the Drude correction of Eq.~\eqref{eq:drude_term}, in our MPA calculations, we find that the Drude tail overlaps with the several inter-band peaks of $Y(\omega)$, without affecting the position of the poles whilst changing their residues (Sec.~IV of \suppinfo), similar to the effect of the CA correction, $Y(\mathbf{q}=0) \sim Y(\mathbf{q_{min}})$, as proposed in Sec.~\ref{sec:intra_analysis}.
In any case, CA is general and independent of the complexity of the frequency structure of the inverse dielectric function $Y$. It works well for Cu, as confirmed by the comparison with the experimental data, and constitutes a very simple procedure.
Despite these considerations, and similarly to the case of Al, the intra-band correction has a small effect on the Cu QP energies, that present differences of the order of 5~meV when computed with and without CA in a $12\times12\times12$ $\mathbf{k}$-grid.

\section{Summary and conclusions}
\label{sec:conclusions}
%

In this work we address the accuracy of the MPA scheme as applied to the full-frequency $GW$ calculation of metals. This approach, previously validated for semiconductors~\cite{Leon2021PRB}, is now applied to metals 
using Al, Na, and Cu as prototype systems.
Also in the case of metals, MPA is shown to deliver results with an accuracy similar to other FF methods at a much lower computational cost. 

After presenting the MPA theoretical framework, we have applied the approach to simple metals and discussed the role of inter- and intra-band contributions to the dielectric functions of bulk Al and Na. 
In order to evaluate the response function and the $GW$ corrections in metals, we have  proposed two simple methods to include the intra-band terms in the inverse dielectric function in the $\mathbf{q} \to 0$ limit: (1) by extrapolating the position of the main pole in $Y_{00}(\mathbf{q},\omega)$, from small $\mathbf{q}$ to $\mathbf{q}=0$, and computing the intra-band pole through the $f$-sum rule of Eq.~\eqref{eq:omega_sum_rule}, which can then be used as an input value in a Drude model to correct $Y_0$. This approach is generalized for a multipole structure of $Y(\mathbf{q},\omega)$ in the case of Cu. And (2) by approximating $Y(\mathbf{q}=0)$ by $Y(\mathbf{q_{min}})$. The second method, here called CA, is simpler and spares the determination of the intra-band frequency.

Both methods significantly accelerate the convergence of the QP energies with respect to the $\mathbf{k}$-point grid. In addition, CA simultaneously corrects all $Y$ matrix elements. CA works equally within PPA, MPA and FF and can be used independently of the dimensionality of the system under study, even if the leading power of series expansion of the inverse dielectric function in the $\mathbf{q} \to 0$ limit depends on dimensionality. In fact, it can be thought of as the most trivial case of a polynomial interpolation (a constant)~\cite{Deslippe2012CPC,Guandalini2022Arxiv}. A similar approach can be applied in situations where the $\mathbf{q} \to 0$ limit of $Y$ (or other many-body operators, such as $W$) is difficult to evaluate. Even if the proposed methodologies were exemplified for three isotropic metals, the extension to non-isotropic systems is straightforward.

Eventually, $GW$ QP corrections for Na, Al and Cu were evaluated, showing an excellent agreement with existing theoretical literature and experimental data, further stressing the accuracy of the proposed approach.
Notably, the case of Cu was discussed with particular detail, since PPA calculations present several drawbacks.
In fact, for Cu, the PPA quasi-particle solutions deviate significantly from the FF results and completely fail to describe the frequency dependence of $\Sigma$ and the spectral function. In contrast, MPA reproduces very accurately the FF results, not only in the tail region that determines the quasi-particles corrections, but in the whole frequency range for both the self-energy and the spectral function.
The frequency representation of the polarizability and the inverse dielectric function present strong oscillations within FF.
In contrast, MPA results are much more stable, leading to a smooth frequency representation of $X$ and $Y$. 

Importantly, the smoother structure of the MPA dielectric function does not necessarily result in a loss of accuracy in the subsequent calculation of the self-energy, the QP energies, and the spectral function. 
In fact, the frequency dependence of $Y$ given by MPA is meaningful and reproduces the main peaks of the experimental energy loss spectra. 
This leads us to conclude that the MPA poles of $Y$  may be seen not only as a mathematical tool, but also as an efficient description of collective excitations, with each pole representing the envelope of a set of single particle transitions.

In conclusion, MPA reproduces well the overall frequency dependence of the polarizability, the inverse dielectric function, the self-energy and the spectral function in metallic systems, and gives results for the quasi-particle energies similar to those obtained within FF  methods. 
Moreover, the favourable computational performance allows for the use of more stringent convergence parameters such as denser $\mathbf{k}$-grids and larger number of bands and polarizability matrices. The use of the proposed intra-band corrections further accelerates the convergence with the $\mathbf{k}$-grid and the accuracy of the final results.

\section*{Acknowledgments}
%
We acknowledge stimulating discussions with Massimo Rontani, Pino D'Amico, Alberto Guandalini and Giacomo Sesti.
This work was partially supported by the MaX -- MAterials design at the eXascale -- European Centre of Excellence, funded by the European Union program H2020-INFRAEDI-2018-1 (Grant No. 824143).
Computational time on the Marconi100 machine at CINECA was provided by the Italian ISCRA program.
%
%
\bibliography{biblio}

\end{document}


\title{Efficient full frequency \texorpdfstring{$GW$}{TEXT} for metals using a multipole approach for the dielectric screening: Supplemental Material}
\author{Dario A. Leon$^{1,2,3}$}
\email{dario.alejandro.leon.valido@nmbu.no}
\author{Andrea Ferretti$^2$}
\author{Daniele Varsano$^{2}$}
\author{Elisa Molinari$^{1,2}$}
\author{Claudia Cardoso$^{2}$}
\affiliation{$^1$ FIM Department, University of Modena \& Reggio Emilia, 41125, Modena (Italy)}
\affiliation{$^2$S3 Centre, Istituto Nanoscienze, CNR, 41125, Modena (Italy)}
\affiliation{$^3$Department of Mechanical Engineering and Technology Management, Norwegian University of Life Sciences, 1430, Ås (Norway)}

\maketitle


%

\section{A simple intra + inter-band model}

In this Section we analyze how two poles in the independent particle response function $Y_0$, 
corresponding to intra- and inter-band transitions, contribute to the structure of its dressed counterpart, $Y$, as a result of the Dyson equation.
%
The $Y_0$ and $Y$ functions are related to the non-interacting and dressed polarizability functions as:
\begin{equation}
    Y_0\equiv v X_0, \qquad
    Y\equiv v X .
\end{equation}
The Dyson equation for the inverse dielectric function is then given by
\begin{equation}
    Y = (1-Y_0)^{-1}Y_0
    \label{eq:dysonY}
\end{equation}
In the following we make use of the $f$-sum rule~\cite{Stefanucci-vanLeeuwen2013book,martin2016book,Lee1994PRB} in the form
%
\begin{equation}
    S_Y({\bf q}) =  \frac{2}{\pi} \int_0^{\infty} d\omega \, \omega \, \text{Im}[Y]({\bf q}, \omega),
\end{equation}
%
where $S$ is computed from the electronic density and the
above expression is valid for each  $(\mathbf{G},\mathbf{G}')$ components of both $Y_0$ and $Y$. In the case $\mathbf{G}=\mathbf{G}'$, one has $S_Y({\bf q}) = \omega^2_\text{pl}$.
%

We then separate $Y_0$ into intra (A) and inter-band (E) contributions in the $\mathbf{q} \to 0$ limit, as
\begin{equation}
    Y_0(\mathbf{q}= 0, \omega) = Y^{A}_0(\omega) +  Y^{E}_0(\omega),
    \label{eq:total_Y}
\end{equation}
resulting in two different terms for the $f$-sum rule:
\begin{equation}
    S_{Y_0}(\mathbf{q}=0) =  S_A + S_E,
\end{equation}
where $S_A$ is the power square of an intra-band frequency, $\Omega_A$, and $S_E$ is obtained from the $f$-sum rule expression for the polarizability $X_0$. When computed from Kohn-Sham (KS) states, 
as derived in Appendix B of Ref.~\cite{Leon2021PRB}, one obtains:
\begin{equation}
    S_E=\sum_n 2 v R^{\text{KS}}_n \Omega^{\text{KS}}_n .
    \label{eq:sum_rule_KS}
\end{equation}

We now analyze the results of Eq.~\eqref{eq:dysonY} in different scenarios by means of a two poles model $y_0(\omega)$. 
The first pole of $y_0(\omega)$ at $\omega=0$ corresponds to a Drude tail resulting from the intra-band transitions. The second broader pole results from the superposition of the inter-band transitions. If we include both single particle contributions in Eq.~\eqref{eq:total_Y}, we obtain:

\begin{equation}
    y_0(\omega) =  \frac{\Omega_A^2}{\omega^2} +  \frac{2 v R_E \Omega_E}{\omega^2 - \Omega_E^2}.
\end{equation}
We consider the inversion of the Dyson equation~\eqref{eq:dysonY} for $y$, disregarding the so called local field effects (i.e.~$y_0$ and $y$ are scalar functions instead of matrices). \\[1pt]

\noindent
Case $S_E\to0$:
\begin{equation}
    y(\omega) =  \frac{\Omega_A^2}{\omega^2 -\Omega_A^2}
\end{equation}
In this case only the intra-band is relevant and we get a pole at the Drude frequency, $\Omega_A$. \\[1pt]

\noindent
Case $S_A\to0$:
\begin{equation}
    y(\omega) =  \frac{2 v R_E \Omega_E}{\omega^2 -\Omega_E^2 - 2 v R_E \Omega_E}
\end{equation}
In this case only the inter-band is relevant and the pole, $\Omega_E$, is right-shifted by a factor $\sqrt{1+2vR_E/\Omega_E}$, which for many materials is close to $1$ (no shift). \\[1pt]

\noindent
General case:
\begin{equation}
    y(\omega) =  \frac{S_1}{\omega^2 -\Omega_1^2} +\frac{S_2}{\omega^2 -\Omega_2^2},
\end{equation}
where the poles are given by the expression:
\begin{multline}
    \Omega_{1,2}^2= \frac{1}{2} \Bigg[ \Omega_E^2 + 2 v R_E \Omega_E + \Omega_A^2  \pm \\
    \sqrt{(\Omega_E^2 + 2 v R_E \Omega_E + \Omega_A^2)^2 - 4 \Omega_E^2 \Omega_A^2} \Bigg],
\label{eq:model_general}
\end{multline}
%
and $S_1$ and $S_2$ are compliant with the $f$-sum rule $S_1+S_2 = S_A+S_E$:
\begin{equation}
    S_{1,2} = \pm \frac{(S_A+S_E)\Omega^2_{1,2}-S_A \Omega_E^2}{\Omega^2_2-\Omega^2_1} .
\end{equation}

The two first cases can be obtained as limiting cases of this general solution. But there are other situations leading to a solution with a single plasmon pole:
\begin{equation}
    y(\omega) =  \frac{2vR_p\Omega_p}{\omega^2 -\Omega_p^2}.
\end{equation}
One case occurs when either $S_1$ or $S_2$ is much larger than the other. A second possibility is when the two poles, $\Omega_1$ and $\Omega_2$, are equal or very close to each other. There are several ways to obtain this: the radicand in Eq.~\eqref{eq:model_general} could be small compared to the terms outside, the scales of the poles could be very different (e.g. $\Omega_E \gg \Omega_A$), $2vR_E/\Omega_E$ could be close to 0 or 1, or the intra- and inter-band poles could be similar, $\Omega_A \approx \Omega_E$.

If $y(\omega)$ has a single pole, the $f$-sum rule for $y$ leads to a simple relationship between the intra-band pole, $\Omega_A$, the inter-band pole, $\Omega_E$ and the plasmon pole, $\Omega_p$:
\begin{equation}
    2 v R_p\Omega_p =\Omega_A^2 + 2vR_E\Omega_E,
    \label{eq:sumrule_model}
\end{equation}
while a similar expression is obtained from the evaluation of $y(0)$:
\begin{equation}
    \Omega_p^2 =\Omega_A^2 + 2vR_E\Omega_E.
\end{equation}
%
From the previous two equations, $v R_p$ and $\Omega_p$ are unequivocally determined, allowing one to discriminate between the intra- and inter-band contributions to the structure of $y$.

For many systems, like the case of Cu addressed in the main manuscript, the total $Y(\omega)$ presents a structure with more than one pole. In this situation the model given in Eq.~\eqref{eq:sumrule_model} can be generalized to obtain the intra-band frequency, again by means of the $f$-sum rule:
\begin{equation}
    \Omega_A^2= \sum 2 v R_p\Omega_p - \sum 2vR_E\Omega_E,
    \label{eq:sumrule_multipole}
\end{equation}
where the sums of $R \Omega$ products run over the pole structures of the full and the inter-band part of $Y(\omega)$.


\section{Analysis of the intra-band limit for different matrix elements}

\begin{figure*}[t]
    \centering
    \includegraphics[width=0.65\textwidth]{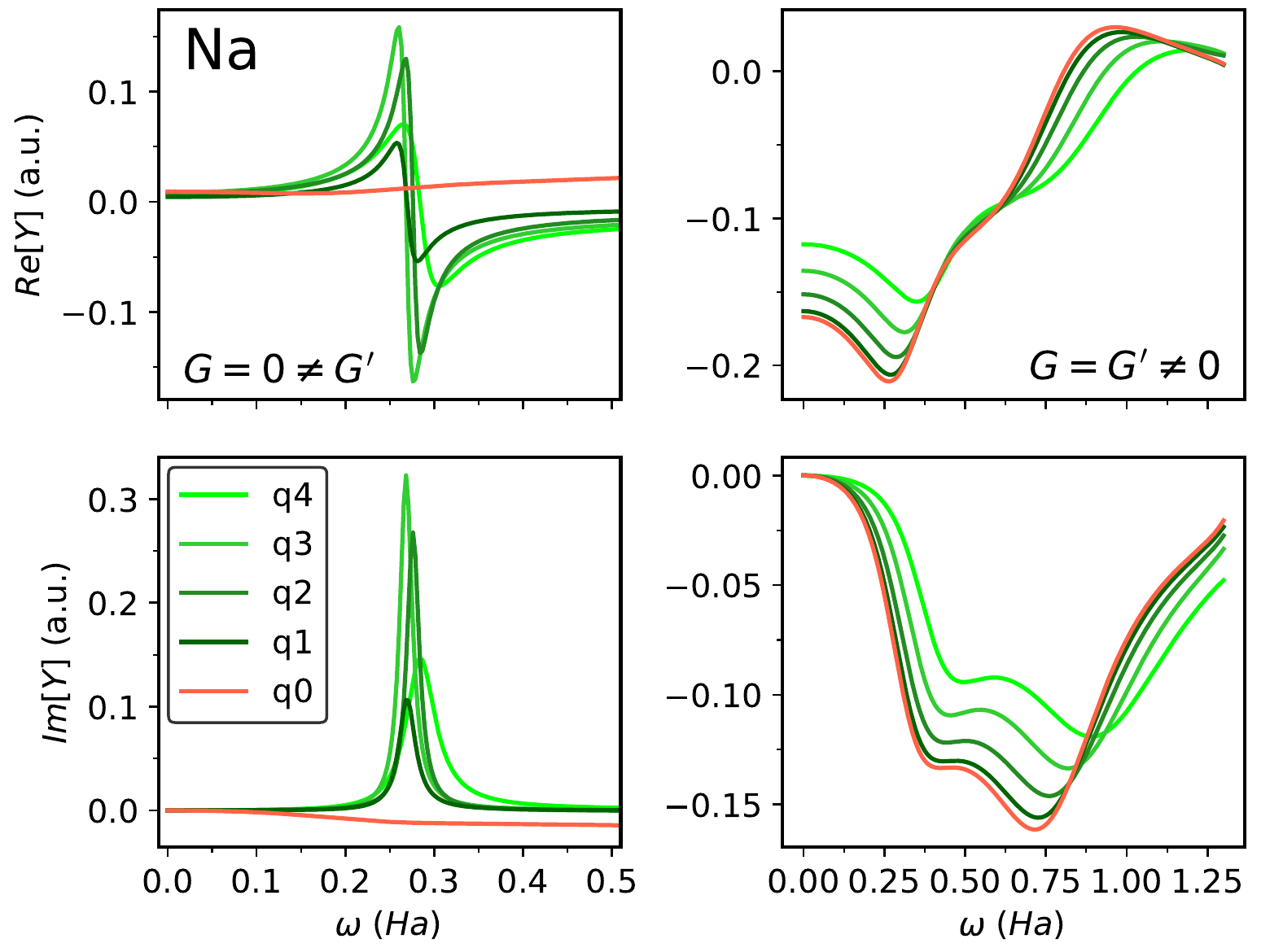}
    \caption{Frequency dependency of $Y$ matrix elements, other than the head ($\mathbf{G}=\mathbf{G'}=0$), computed with MPA for different $\mathbf{q}$ vectors of modulus \DAL{$q\equiv |\mathbf{q}|$} tending to 0, for Na. For $q=0$ (orange curves) the intra-band term is not included. The chosen values of $q_0$--$q_4$ correspond to $q_n=\frac{n}{8}$ in units of $2\pi/a$, with $a$ the lattice parameter of Na.}
    \label{fig:X_Na}
\end{figure*}

In the main text we have shown  that the  $Y_{\mathbf{G}=\mathbf{G'}=0}$ curves with finite $\mathbf{q}$, computed for Al and Na, change smoothly with decreasing $\mathbf{q}$. The curves present a pole, $\Omega_p(\mathbf{q})$, of decreasing energy and increasing amplitude and both the real and imaginary part of this pole can be easily extrapolated to $\mathbf{q}=0$, by means of the Lindhard plasmon dispersion.
Here we have considered other Na $Y$ matrix elements ($Y_{\textbf{G}=\textbf{G'}\neq 0}$ and $Y_{\textbf{G}\neq \textbf{G'}}$) with the same not particularly dense grid of $8\times8\times8$ {\bf k}-points.

In Fig.~\ref{fig:X_Na} we show similar plots for 
wing ($Y_{\textbf{G}=0\neq \textbf{G'}}$ and $Y_{\textbf{G}\neq 0 =\textbf{G'}}$) and diagonal ($Y_{\textbf{G}=\textbf{G'}\neq 0}$) elements, where the orange curves contain only inter-band terms.
For some of these elements the evolution of the $Y$ curves when $ \mathbf{q}\to 0$ is less smooth than for the head, $Y_{\mathbf{G}=\mathbf{G'}=0}$, and the position of the poles not always changes monotonously, thus the extrapolation with a simple analytical form would require a denser $\mathbf{k}$-point grid depending on the specific matrix element.
However, as can be seen from the comparison between green and orange curves, intra-band transitions are more important for the wings than for diagonal elements.
Moreover, even if the constant dielectric function approximation (CA) scheme described in the main manuscript, i.e, $Y(\mathbf{q}=0)\sim Y(\mathbf{q_{min}})$, has limited accuracy for some of these matrix elements, it still improves the overall $Y$ matrix.

\section{Frequency representability of copper}

\begin{figure}
    \centering
    \includegraphics[width=0.35\textwidth]{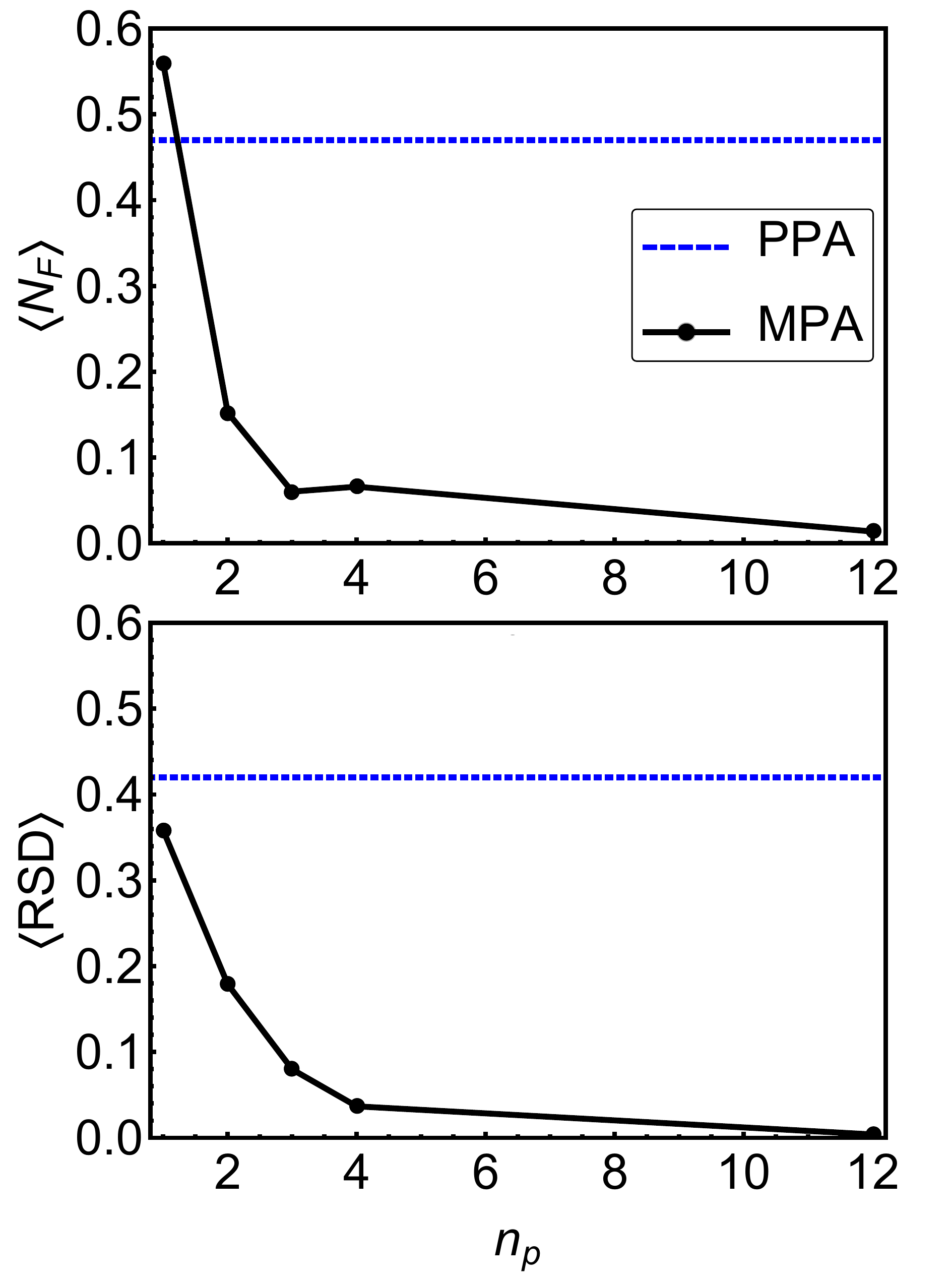}
    \caption{Values calculated for Cu of (left) the mean number of matrix elements,  $\langle N_F \rangle$, for which the position of the poles was corrected according to Eq.~\eqref{eq:MPcond1} for MPA and Eq.~\eqref{eq:PPcond} for PPA.}
    \label{fig:Cu_failure}
\end{figure}

As mentioned in the main manuscript, the nonlinear interpolation of the polarizability, $X$, or the inverse dielectric function, $Y$, may produce non-physical poles corresponding to the so-called {\it unfulfilled modes}. This is usually solved by reassigning the values of the poles. The condition used to identify unfulfilled modes in the GN-PPA scheme~\cite{Leon2021PRB} is the following:
\begin{equation}
    \text{Re}\left[ \frac{ Y_{\mathbf{G}\mathbf{G}'}(\mathbf{q},0)}{Y_{\mathbf{G}\mathbf{G}'}(\mathbf{q},i \varpi_\text{pl})} -1 \right] <0, 
    \label{eq:PPcond}
\end{equation}
where the position of the pole is set to $\Omega_{\text{fail}}^{\text{GN}}=1$~Ha in case of failure. 

The MPA scheme uses a generalized condition that avoids reassigning the poles with a constant value~\cite{Leon2021PRB}:
\begin{equation}
 \Omega_n = \left\{
    \begin{aligned}
     &\sqrt{\Omega_n^2}, \qquad \qquad \text{Re}\left[ \Omega_n^2 \right] \geq 0 \\
     &\sqrt{-(\Omega_n^2)^*}, \qquad \text{Re}\left[ \Omega_n^2 \right] < 0 
    \end{aligned}
 \right.
 \label{eq:MPcond1}
\end{equation}

It is possible  to quantify the representability error related to this reassignment by computing the mean number of corrected $X$ matrix elements, $\langle N_F \rangle$, and an average relative standard deviation of the extrapolated $X$ with respect to its sampling points, $\langle RSD \rangle$, as defined in Ref.~\cite{Leon2021PRB}. In Fig.~\ref{fig:Cu_failure} we show how these two quantities evolve when increasing the number of poles used in the description of $X$ and $Y$. 

When applying the PPA, we found that $48\%$ of the polarizability matrix elements fail the plasmon pole condition  \eqref{eq:PPcond}. These results have an averaged relative deviation of $\langle RSD \rangle=0.42$. These large values lead to quasi-particle solutions very different from FF, as described in the main text. The MPA scheme, with only one pole, still presents a larger percentage of corrected poles but considerably improves the representability with respect to PPA, lowering the average deviation to $\langle RSD \rangle=0.35$. The reason for the larger number of corrected poles is the use of a different sampling. As mentioned in the main manuscript, in the case of MPA with one pole the frequency at the origin of coordinates is shifted along the imaginary axis, which helps to reduce the numerical instabilities found with the PPA sampling~\cite{Marini2002PRL}. However, a significant improvement happens only by increasing the number of poles, as illustrated in Fig.~\ref{fig:Cu_failure}, evidencing the complexity of the frequency structure of the polarizability of Cu and the efficiency of the multipole approximation in its description.


\section{Standard Drude correction for the intra-band of copper}

\begin{figure}
    \centering
    \includegraphics[width=0.4\textwidth]{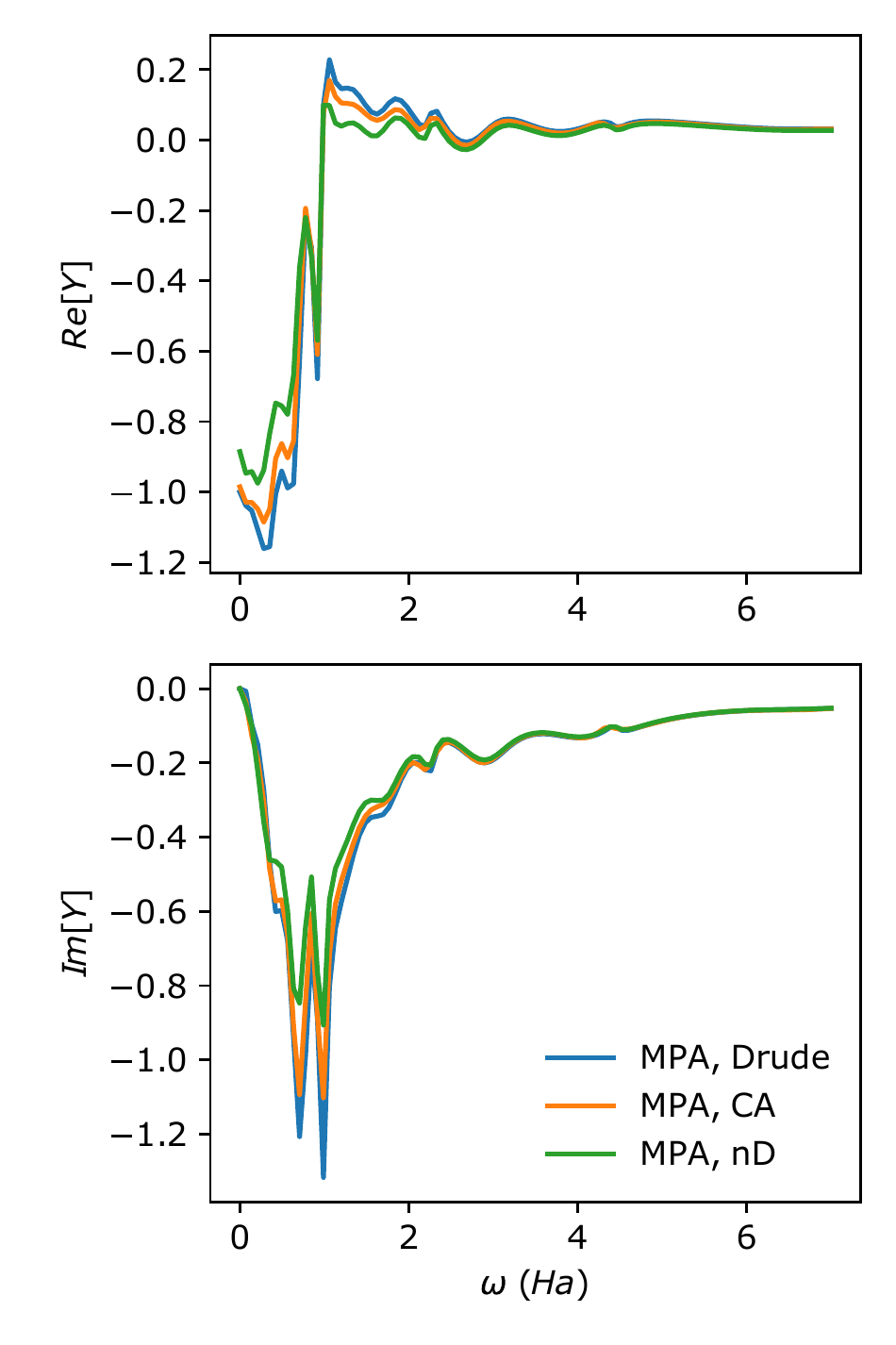}
    \caption{Frequency dependence of $Y(\mathbf{q}=0)$ computed for Cu with three different methods: without any intra-band correction (nD), with the CA correction and with the Drude model using as input the intra-band frequency determined in Ref.~\cite{Marini2001PRB}.}
    \label{fig:Cu_CA_Drude_nD}
\end{figure}

In Fig.~\ref{fig:Cu_CA_Drude_nD} we show the frequency dependence of $Y$ computed for Cu within MPA comparing three different methods to include the intra-band corrections in the limit $q\to0$. We show, $Y$ computed without any intra-band correction (nD), with the CA method and the Drude model. For the latter, we considered an intra-band frequency of 9.27~eV, as determined in Ref.~\cite{Marini2001PRB}. We see that the Drude model and CA give similar results, renormalizing the intensity of the peaks without shifting them in the real frequency axis.

\section{Convergence of $GW$ parameters for copper}

\begin{figure*}
    \centering
    \includegraphics[width=1\textwidth]{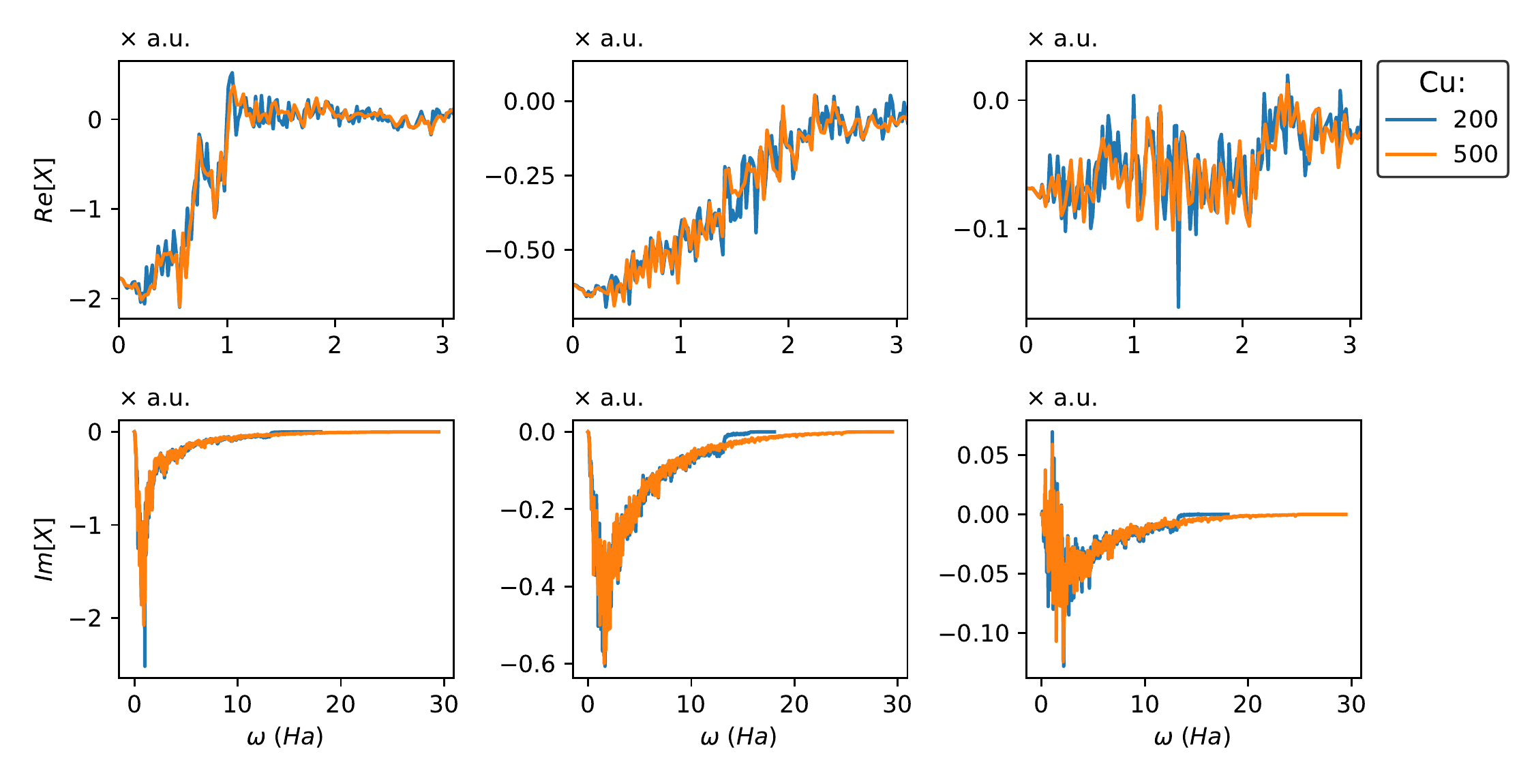}
    \caption{Selected Cu $X$ matrix elements computed within FF with different number of bands (200 and 500). In order to show both, the overall and the detailed behavior of the polarizability, we plotted the imaginary part of $X$ in the bottom panels in the full frequency interval determined by the number of bands, while the real part of $X$ are plotted in the top panels in a zoomed region. A similar scheme from Ref.~\cite{Leon2021PRB} is used for the units of X, where we omitted the specific scales in order to avoid confusion with the zoomed plots.} 
    \label{fig:XpolFF_Cu}
\end{figure*}

\begin{figure*}
    \centering
    \includegraphics[width=0.55\textwidth]{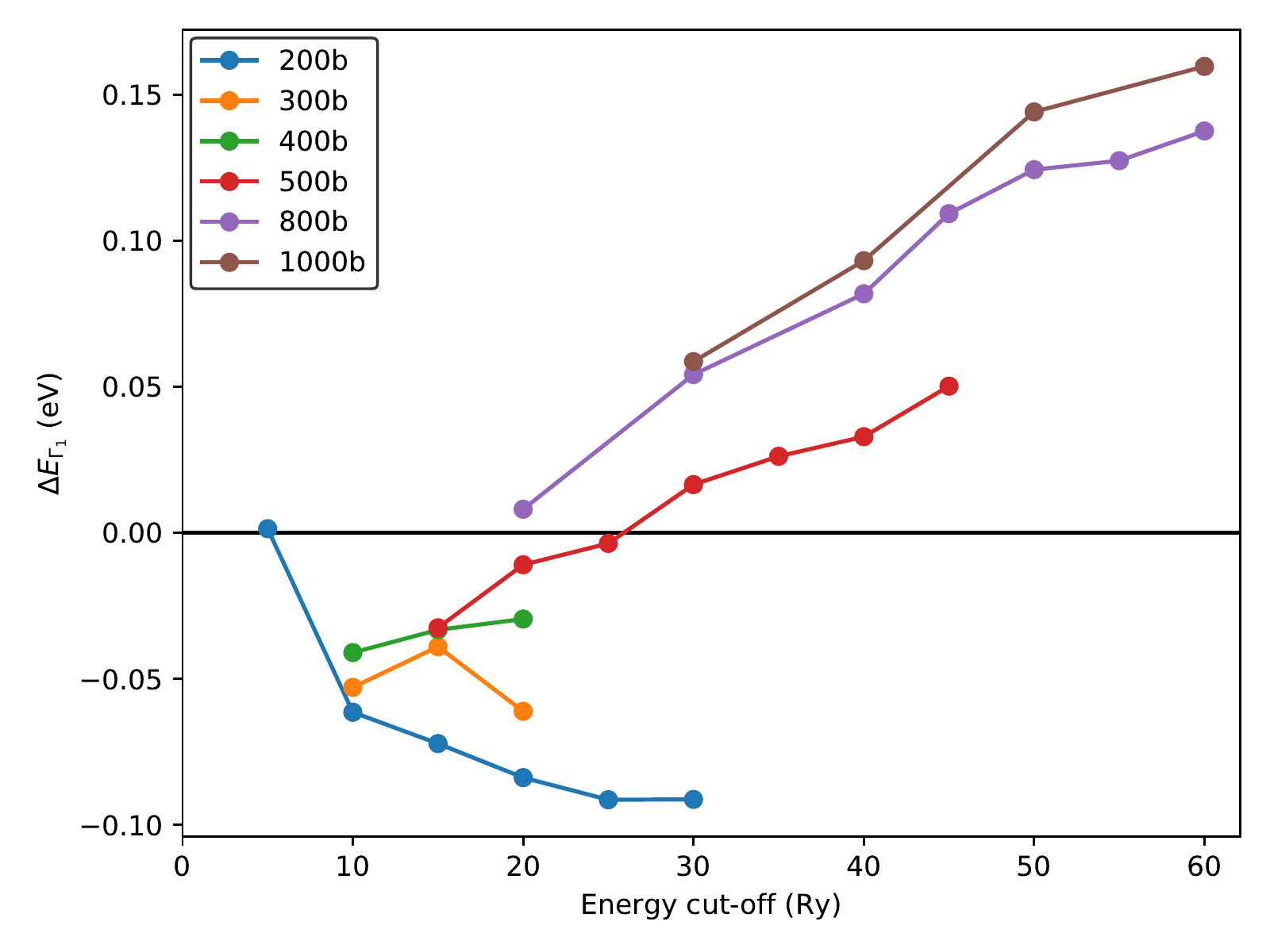}
    \caption{$G_0W_0$ correction to the PBE quasi-particle energy of the $\Gamma_1$ state of Cu. Convergence with respect to the number of bands and the cut-off energy parameters used to build the polarizability matrix, $X$. The curves correspond to calculations with different fixed number of bands.}
    \label{fig:Cu_convergence}
\end{figure*}

As mentioned in the main text, the $GW$ convergence is very challenging for the case of Cu. In Fig.~\ref{fig:XpolFF_Cu} we illustrate how the energy range of the transitions observed in $X$ rapidly increases with respect to the number of bands included in the calculation. 
The increasing number of bands 
results in changes in the details of the frequency structure of $X$, whose description requires, in the FF-RA scheme, a large number of frequency points.

On the other hand, there is a complex relationship between the number of bands and the plane-wave energy cut-off used to build the polarizability matrix. As shown in Fig.~\ref{fig:Cu_convergence}, there is a change of monotony around 15 Ry when increasing the number of bands from 200 to 500 and smaller oscillation continue at higher cut-off energies, hindering the possibility to extrapolate the converged QPs. For cut-off energies and number of bands larger than 25~Ry and 500 respectively, the correction changes sign, correcting the DFT in the right direction with respect to the experimental data shown in the Table. II of the main manuscript.

%
%
\bibliography{biblio}